\documentclass[%
 reprint,
 amsmath,amssymb,
 aps,
prb,
]{revtex4-2}

\usepackage{graphicx, color}% Include figure files
\usepackage{dcolumn}% Align table columns on decimal point
\usepackage{bm}% bold math
\usepackage{hyperref}% add hypertext capabilities
\hypersetup{
 citecolor=blue,
 colorlinks = true,
 urlcolor = blue
}
\usepackage[normalem]{ulem}%取り消し線用
\usepackage{physics}
\usepackage{ulem}
\usepackage{xcolor}
\usepackage{cancel} % 取り消し線を引くためのパッケージ
\allowdisplaybreaks[1]

\begin{document}

\title{Band curvature effects on quantum transport of spin-1 chiral fermion systems}% Force line breaks with \\

\author{Risako Kikuchi}
\affiliation{Department of Physics, Nagoya University, Nagoya 464-8602, Japan}
\author{Ai Yamakage}
\affiliation{Department of Physics, Nagoya University, Nagoya 464-8602, Japan}

\date{\today}

\begin{abstract}
We theoretically investigate the quantum transport properties of three-dimensional spin-1 chiral fermion systems with a curved trivial band. 
In the multiband system with two distinct characters--a linear Dirac band and a quadratic trivial band--the hybridization induced by impurity effects leads to pronounced energy and temperature dependences in the electrical conductivity.
We show that the conductivity is suppressed by the trivial band in the low-energy regime near the threefold degenerate point and enhanced in the band-crossing point of the Dirac and trivial bands. These results are derived using the self-consistent Born approximation within the framework of linear response theory.
\end{abstract}

\maketitle

\section{\label{sec:level1}Introduction}
The discovery of topological semimetals with multiple energy bands has opened a new frontier in the field of topological physics \cite{Ma2012, Beyond2016}. 
Topological semimetals are characterized by the crossing of multiple energy bands, resulting in distinctive physical properties \cite{Lv2021}. 
Among these, the threefold degeneracy, which consists of two linear bands and one nearly flat ``trivial band," is expected to exhibit unique physical phenomena arising not only from the singular nature of the band crossing but also due to the presence of the trivial band. 
The threefold crossing is known as a spin-1 chiral fermion. 
The three-dimensional spin-1 chiral fermion systems are theoretically predicted to emerge in chiral crystals and have been experimentally observed in materials \cite{Tang2017-kk, Chang2017, Pshenay-Severin_2018, Takane2019, Rao2019-ts, Sanchez2019-by, Mozaffari2020, Robredo_2024} such as the B20 family, including CoSi. 
Recently, transport phenomena \cite{Tang21, Kikuchi2022, Lien2023, Kikuchi2023, Selma2024, Nakazawa2024-nk}, quantum oscillations \cite{Wu_2019, Xu2019, Petrova2023, Leeb2023, Huber2023}, optical response \cite{Flicker18, Sanchez-Martinez2019-ek, Li2019, Habe2019-ss, Maulana2020, Chang2020, Dylan2020, Xu2020-zb, Le2020, Ni2020-ze, Ni2021-eo, Rees2021, Kaushik2021-hx, Dey2022-th, Lu2022, Hsiu2023, Yang2023}, superconductivity \cite{Mandal2021-ic, Gao2022}, and quantum phenomena originating from topological structures \cite{Yuan2019-hz, Huber2022, hsu2022disorder, Balduini2024} of the spin-1 chiral fermion materials have been studied, revealing exotic quantum phenomena. 

Quantum transport in topological semimetals, including two-band systems such as Dirac and Weyl semimetals \cite{Noro2010-cm, 0minato2014, Kobayashi2014, Nandkishore2014-vz, Ominato2015-um, Ominato2016-jl}, is well known to be highly sensitive to impurity scattering \cite{hsu2022disorder, Kikuchi2022, Kikuchi2023, Leeb2023, Selma2024}.
In particular, the importance of incorporating impurity scattering through the self-consistent effect has been extensively discussed \cite{Noro2010-cm, 0minato2014, Ominato2015-um, Kikuchi2022, Kikuchi2023, Leeb2023}.
We previously showed the transport properties of a three-dimensional spin-1 chiral fermion system at absolute zero, assuming a perfectly flat trivial band \cite{Kikuchi2022, Kikuchi2023}. 
Our results revealed a suppression of electrical conductivity due to the influence of the flat trivial band.
However, in realistic spin-1 chiral fermion systems, the trivial band is not completely flat but exhibits some curvature. 
In order to more accurately capture the transport phenomena in real materials, it is necessary to consider the case of a curved trivial band and to reproduce the transport properties at finite temperatures.

In this study, we investigate the quantum transport phenomena in spin-1 chiral fermion systems, focusing specifically on the role of the curvature of the trivial band. 
By employing the self-consistent Born approximation (SCBA) within the framework of linear response theory, we demonstrate that the electrical conductivity is notably suppressed near the band edge of the trivial band.
This distinctive behavior results directly from the multiband nature of the system and is attributed to the effect of the band mixing by the impurity potential.
Another multiband effect for the electrical conductivity is the enhancement at the twofold-degenerate point.
Additionally, the conductivity strongly depends on the temperature: metallic-like in the high-energy region and insulating-like in the low-energy region. 
These insights deepen our understanding of the interplay between multiband structure and quantum transport.

The structure of this paper is as follows.
Section~\ref{model} introduces the model of spin-1 chiral fermion systems. 
Section~\ref{linear1} describes the method used to calculate the density of states and electrical conductivity within the SCBA in linear response theory, followed by the presentation of results in Sec.~\ref{linear2}. 
In Sec.~\ref{Boltzmann}, we derive the conductivity using the Boltzmann equation and compare the results with those obtained by the SCBA. 
Section~\ref{discussion} discusses the potential for experimentally observing the theoretical results of this study and the specific parameter values relevant to such observations.
Finally, Sec.~\ref{conclusion} summarizes this work.

\section{model}
\label{model}

\subsection{Spin-1 chiral fermions}
We consider a general three-dimensional spin-1 chiral fermion system that is isotropic and time-reversal invariant, described by \cite{Mandal2021-ic}
\begin{align}
\hat{\mathcal{H}}=\hbar v\hat{\bm{S}}\cdot \bm{k}
+ c'[(\hat{\bm{S}}\cdot \bm{k})^2
- b' k^2\hat{S}_0],
\end{align}
where $\bm k$ is the electron wavenumber and $v$ is the Fermi velocity. $\bm{S}=(\hat{S}_{x},\hat{S}_{y},\hat{S}_{z})$ are the representation matrices for spin-1 \cite{Beyond2016}: 
\begin{align}
\hat{S}_{x} &=
\pmqty{
0 & i & 0\\
-i & 0 & 0\\
0 & 0 & 0
},
\\
\hat{S}_{y}&=
\pmqty{
0 & 0 & -i\\
0 & 0 & 0\\
i & 0 & 0
},
 \\
\hat{S}_{z} &=
\begin{pmatrix}
0 & 0 & 0\\
0 & 0 & i\\
0 & -i & 0\\
\end{pmatrix}.
\end{align}
The eigenenergy $\epsilon_{\lambda,\bm{k}}$ is given by 
$
\epsilon_{\text{c},\bm{k}}
=\hbar v k + c' (1-b') k^2, 
\
\epsilon_{\text{t},\bm{k}}
=-c' b' k^2,
\
\epsilon_{\text{v},\bm{k}}
=-\hbar v k + c' (1-b') k^2,
$
where $\lambda$ is the label for the conduction band ($\lambda = \mathrm c$), the trivial band ($\lambda = \mathrm t$), and the valence band ($\lambda = \mathrm v$).
For the purpose of examining the effects of the curvature of the trivial band, the parameter is fixed as $b'=1$. 
Under this condition, two of the three energy bands display entirely linear dispersion, with the third being a trivial band that has curvature as
\begin{align}
\epsilon_{\text{c},\bm{k}}
&=\hbar v k,
\\
\epsilon_{\text{t},\bm{k}}
&=-c' k^2,\\
\epsilon_{\text{v},\bm{k}}
&=-\hbar v k.
\end{align}
{The eigenstates $\vb*{\xi}_{\lambda, \vb*{k}}$ of the spin-1 fermion system are written as
\begin{align}
 \vb*{\xi}_{+1, \vb*{k}} 
&=\frac{1}{\sqrt{2}k\sqrt{k_x^2+k_z^2}}\left( \begin{array}{c} k_yk_z-ikk_x \\ -k_x^2-k_z^2 \\ k_xk_y+ikk_z \end{array} \right),\label{eigenstates1}\\
\vb*{\xi}_{t, \vb*{k}}
&=\frac{1}{k}\left( \begin{array}{c} k_z \\ k_y \\ k_x \end{array} \right),\\
\vb*{\xi}_{-1, \vb*{k}}
&=\frac{1}{\sqrt{2}k\sqrt{k_x^2+k_z^2}}\left( \begin{array}{c} k_yk_z+ikk_x \\ -k_x^2-k_z^2 \\ k_xk_y-ikk_z \end{array} \right).\label{eigenstates2}
\end{align}}

%------------------------------------------------------------%
\begin{figure}
\includegraphics[width=5cm]{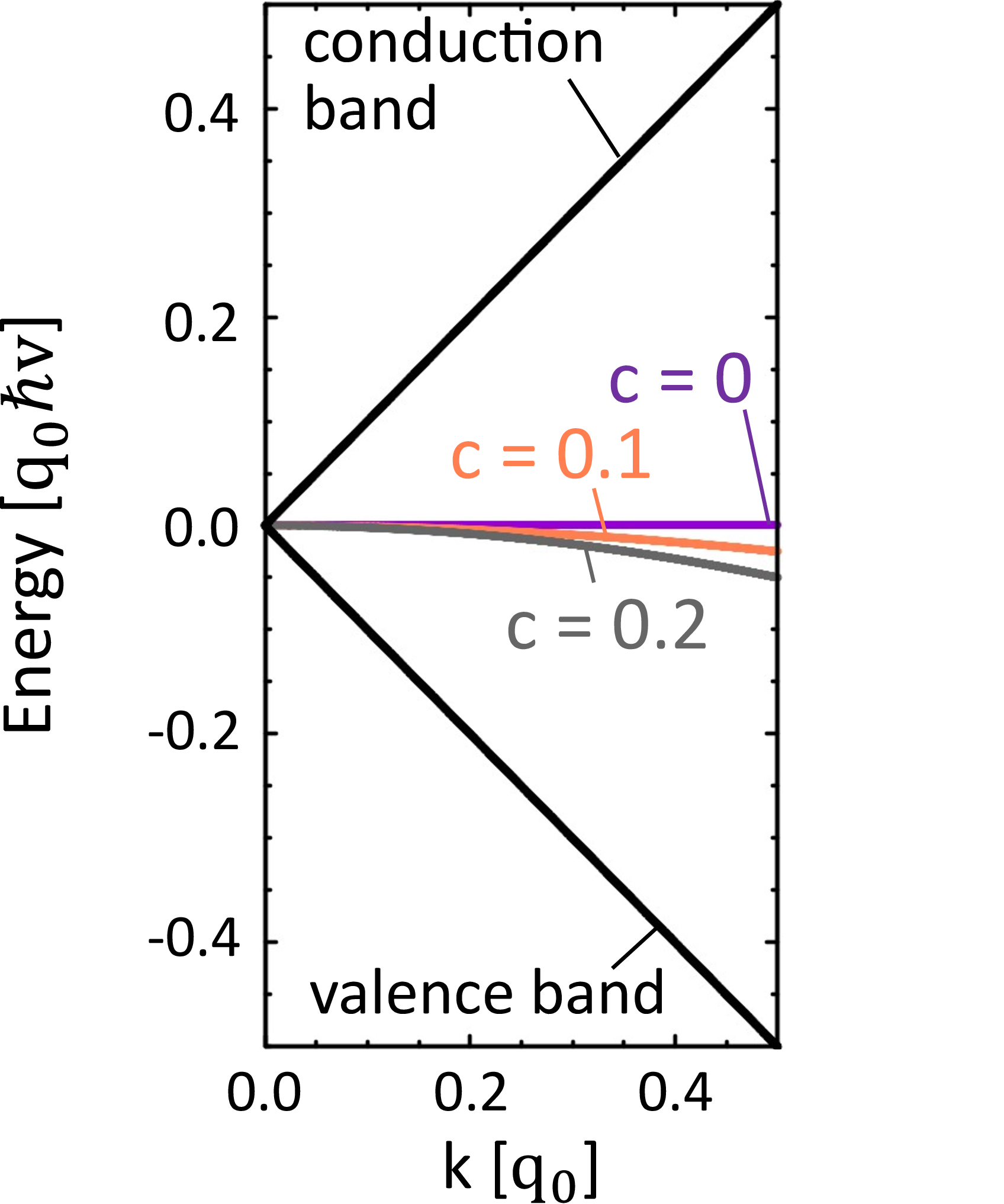}
\caption{(Color online) Energy band of the spin-1 chiral fermion system for $c=0$ (purple), $c=0.1$ (orange), and $c=0.2$ (gray).}
\label{fig_band} 
\end{figure}
%------------------------------------------------------------%

\subsection{Disorder}
We assume Gaussian impurity potentials, which represent finite-range disorders.
The Gaussian potential is defined by
\begin{align}
\label{gauss}
U(\bm{r}) = \frac{\pm u_0}{(\sqrt{\pi}d_0)^3}\exp(-\frac{r^2}{d_0^2}),
\end{align}
where $d_0$ is the characteristic length scale and $\pm u_0$ is the strength of the impurity potential. 
The sign $\pm$ means to assume that the numbers of positive and negative valued impurities are the same, so the Fermi level is fixed, irrelevant to the impurity concentration.
The Fourier transform is obtained to be
\begin{align}
 u(\bm{k}) = 
  \int d^3 r 
  \,
  e^{-i \boldsymbol{k} \cdot \boldsymbol{r}} U(\boldsymbol{r})
 =\pm u_0\exp(-\frac{k^2}{q_0^2}),
 \label{u_Gauss}
\end{align}
with $q_0 = 2/d_0$. 

We introduce a parameter that quantifies the scattering strength: 
\begin{eqnarray} 
W = \frac{q_0n_{\text{i}}u_0^2}{\hbar^2 v^2}, 
\end{eqnarray} 
where $n_{\text{i}}$ represents the number of scatterers per unit volume.
We also define the dimensionless parameter characterizing the curvature of the trivial band:
\begin{align}
c=\frac{c' q_0}{\hbar v}.
\end{align}

\subsection{Density of states in the clean limit}
The energy-band structure of the spin-1 fermion system is shown in Fig.~\ref{fig_band}, depending on the value of $c$.
The density of states per unit volume $D_{0}(\epsilon)$ in the clean limit is given by
%for energy band $\lambda= \mathrm{c}, \mathrm{t}, \mathrm{v}$ is given by
%\begin{align}
  %&D_{0,\mathrm c}(\epsilon)=\frac{\epsilon^2}{2\pi^2(\hbar v)^3}\theta(\epsilon),
%  \\
%  &D_{0,\mathrm t}(\epsilon)=\frac{q_0}{4\pi^2 |c| \hbar v}\sqrt{\frac{q_0\epsilon}{|c|\hbar v}}\theta(c\epsilon), 
%  \\
%  &D_{0,\mathrm v}(\epsilon)=\frac{\epsilon^2}{2\pi^2(\hbar v)^3}\theta(-\epsilon).
%\end{align}
%By summing these contributions, the total clean density of states for the spin-1 fermion system is 
\begin{align}
D_{0}(\epsilon)&=
\frac{\epsilon^2}{2\pi^2(\hbar v)^3}+\frac{1}{4\pi^2 |c'|}
\sqrt{-\frac{\epsilon}{c'}}\theta(-c'\epsilon) \label{eq_clean}.
\end{align}
The second term comes from the trivial band. 
%When the trivial band bends in $\mu<0$ (for $c>0$), the presence of the trivial band leads to an increase in the density of states in $\mu<0$.

%------------------------------------------------------------%
%\begin{figure}
%\includegraphics[width=8cm]{Fig-clean}
%\caption{(Color online) Density of states (DOS) in the clean limit ($W=0$) for $c=0$ (purple solid line), $c=0.1$ (orange dashed line), and $c=0.2$ (gray dotted line). }
%\label{fig_clean} 
%\end{figure}
%------------------------------------------------------------%

\section{\label{linear1}Formulation of self-consistent Born approximation}
We then proceed to calculate the density of states (DOS) and electrical conductivity using linear response theory with SCBA. 

\subsection{Formulation}
Assuming a uniform random distribution of impurities, the impurity-averaged Green's function is expressed as follows:
\begin{widetext}
\begin{align}
%&
\hat{G}(\bm{k},\epsilon +is0) =
%\nonumber\\ & 
\frac{1}{\qty[\epsilon+(c\hbar v/q_0)k^2] \hat{S}_0-\hbar vk\hat{\bm{S}}\cdot\bm{n}-(c\hbar v/q_0)k^2(\hat{\bm{S}}\cdot \bm{n})^2-\hat{\Sigma}(\bm{k},\epsilon+is0)}, 
\label{green function}
\end{align}
\end{widetext}
where $\bm{n} = \bm{k}/k$ is the unit vector and $\hat{S}_0$ represents the identity matrix.
The sign $s$ denotes the retarded ($s=1$) and advanced ($s=-1$) Green's functions.
The self-consistent equation for the self-energy is given by 
\begin{align}
\hat{\Sigma}(\bm{k},\epsilon+is0)
 = \int\frac{d\bm{k'}}{(2\pi)^3}n_{\text{i}}|u(\bm{k}-\bm{k'})|^2\hat{G}(\bm{k'},\epsilon+is0).
 \label{self energy}
\end{align}
The DOS per unit volume is calculated as
\begin{align}
D(\epsilon) = -\frac{1}{\pi}\Im\int\frac{d\bm{k}}{(2\pi)^3}\Tr\hat{G}(\bm{k},\epsilon+i0).
\label{dos}
\end{align}
The conductivity, as given by the Kubo formula, can be expressed as
\begin{align}
\sigma(\mu) &= -\frac{\hbar e^2 v^2}{4\pi}\sum _{s,s'=\pm1}ss'
\int\frac{d\bm{k'}d\epsilon}{(2\pi)^3}
\qty(
    -\frac{\partial f}{\partial\epsilon}
    )
\notag\\& \quad \times
\text{Tr} \biggl[
%\textcolor{red}
{\frac{\hat{v}_x}{v}}\hat{G}(\bm{k'},\epsilon+is0)
%\nonumber\\& \quad\times
\hat{J}_x(\bm{k'},\epsilon+is0,\epsilon+is'0)
\notag\\&\hspace{4em}\times
\hat{G}(\bm{k'},\epsilon+is'0)
\biggr],\label{conductivity}
\end{align}
where {$f(\epsilon)=[e^{(\epsilon-\mu)/(k_{\mathrm{B}}T)}+1]^{-1}$ is the Fermi-Dirac distribution function and}
\begin{align}
 \hat v_x
 &=\frac{1}{\hbar}\frac{\partial \hat{\mathcal{H}}}{\partial k_x}\nonumber\\
 &= v \hat S_x
 - 2 \frac{cv}{q_0} k_x \hat S_0
 + 
 \frac{cv}{q_0}
 (\hat{\boldsymbol{S}} \cdot \boldsymbol{k}) \hat S_x
 +
 \frac{cv}{q_0}
 \hat S_x
 (\hat{\boldsymbol{S}} \cdot \boldsymbol{k}), 
\end{align}
denotes the velocity matrix along the $x$ direction and $\hat{J}_x(\bm{k},\epsilon,\epsilon')$ represents the corresponding current vertex function, including the vertex correction 
determined as the solution to the following Bethe-Salpeter equation: 
\begin{align}
\hat{J}_x(\bm{k},\epsilon,\epsilon') &=
{\frac{\hat{v}_x}{v}} + \int\frac{d\bm{k'}}{(2\pi)^3}n_{\text{i}}|u(\bm{k}-\bm{k'})|^2\hat{G}(\bm{k'},\epsilon)\nonumber\\
&\quad\times
\hat{J}_x(\bm{k'},\epsilon,\epsilon')\hat{G}(\bm{k'},\epsilon')\label{Bethe}
\nonumber\\
& \hspace{-2em} =
\hat{S}_x 
-\frac{2c k}{q_0}n_x\hat{S}_0
+
\frac{c k}{q_0}(\hat{\bm{S}}\cdot \bm{n})\hat{S}_x
+
\frac{c k}{q_0}
\hat{S}_x
(\hat{\bm{S}}\cdot \bm{n})
\notag\\&\quad
+ \int\frac{d\bm{k'}}{(2\pi)^3}
n_{\text{i}}|u(\bm{k}-\bm{k'})|^2
\nonumber\\
&\qquad\times
\hat{G}(\bm{k'},\epsilon)\hat{J}_x(\bm{k'},\epsilon,\epsilon')\hat{G}(\bm{k'},\epsilon').
\end{align}
The detailed calculations are provided in Appendix \ref{calculations}.

\subsection{Intraband and interband contributions}

For a thorough understanding of physical quantities in multiorbital systems, it is useful to decompose them into intraband and interband components. These components can be obtained by diagonalizing the Green's function matrix as follows:
\begin{align}
\hat{U}^{\dagger}\hat{G}(\bm{k},\epsilon+is0)\hat{U}=
\begin{pmatrix}
G^s_c & 0 & 0\\
0 & G^s_0 & 0\\
0 & 0 & G^s_v\\
\end{pmatrix}.
\end{align}
where the subscripts $c$, $0$, and $v$ represent the conduction, trivial, and valence bands in the band basis, respectively. In this basis, the velocities $\hat{v}_{x}$ and $\hat{J_{x}}$ can be written as
\begin{align}
\frac{1}{v}\hat{U}^{\dagger}\hat{v}_{x}\hat{U}=
\begin{pmatrix}
v_{cc} & v_{c0} & 0\\
v_{0c} & v_{00} & v_{0v}\\
0 & v_{v0} & v_{vv}\\
\end{pmatrix},
\end{align}
and
\begin{align}
\hat{U}^{\dagger}\hat{J_{x}}(k, \epsilon+is0, \epsilon+is'0)\hat{U}=
\begin{pmatrix}
J^{ss'}_{cc} & J^{ss'}_{c0} & 0\\
J^{ss'}_{0c} & J^{ss'}_{00} & J^{ss'}_{0v}\\
0 & J^{ss'}_{v0} & J^{ss'}_{vv}\\
\end{pmatrix}.
\end{align}
We decompose the DOS into the Dirac-cone and the trivial-band terms as
\begin{align}
D_{\text{Dirac}}(\epsilon)
&= -s\frac{1}{\pi}\Im\int\frac{d\bm{k}}{(2\pi)^3}(G^s_c+G^s_v)\label{dosdirac},
\end{align}
and
\begin{align}
D_{\text{trivial}}(\epsilon)&=-s\frac{1}{\pi}\Im\int\frac{d\bm{k}}{(2\pi)^3}G^s_0.
\label{dosflat}
\end{align}

Additionally, we decompose the conductivity into intraband and interband terms as follows: 
\begin{align}
\sigma_{\text{D-D}}(\mu) &= -\frac{\hbar e^2 v^2}{4\pi}\sum _{s,s'=\pm1}ss' \int d\epsilon{\qty(
    -\frac{\partial f}{\partial\epsilon}
    )}\int\frac{d\bm{k'}}{(2\pi)^3}
\notag\\&\quad\times
\bigl(v_{cc}G^s_c J^{ss'}_{cc}G^{s'}_c
%\nonumber\\&\qquad
+v_{vv} G^s_v J^{ss'}_{vv} G^{s'}_v  \bigr).
\label{intra}
\end{align}
The interband term between the Dirac cone and the trivial band is defined by
\begin{align}
\sigma_{\text{D-t}}(\mu) &= -\frac{\hbar e^2 v^2}{4\pi}\sum _{s,s'=\pm1}ss'
\int d\epsilon{\qty(
    -\frac{\partial f}{\partial\epsilon}
    )}\int\frac{d\bm{k'}}{(2\pi)^3}
\notag\\&\quad\times
\bigl(v_{0c} G^s_c J^{ss'}_{c0} G^{s'}_0\nonumber\\
&\qquad
+v_{c0} G^s_0 J^{ss'}_{0c} G^{s'}_c +v_{0v} G^s_v J^{ss'}_{v0} G^{s'}_0 \nonumber\\&\qquad
+v_{v0} G^s_0 J^{ss'}_{0v} G^{s'}_v \bigr).
\label{inter}
\end{align}
The intraband term of the trivial band is given by
\begin{align}
\sigma_{\text{t-t}}(\mu) 
&= -\frac{\hbar e^2 v^2}{4\pi}\sum _{s,s'=\pm1}ss'\int d\epsilon{\qty(
    -\frac{\partial f}{\partial\epsilon}
    )}\int\frac{d\bm{k'}}{(2\pi)^3}
\notag\\&\qquad \times
v_{00}G^s_0 J^{ss'}_{00}G^{s'}_0.
\label{intra}
\end{align}
%

%\begin{figure*}
%	\includegraphics[width=17.5cm]{band}
%	\caption{(Color online) The conductivity from the intraband transition of the Dirac cone (black line) and from the interband transition between the Dirac cone and the flat band (red line) for $\alpha=0.02$ derived by the SCBA. (a) delta function potential, (b) the Coulomb potential, $k_{\text{c}}=q_0$ and (c) the Coulomb potential, $k_{\text{c}}=3q_0$}
%	\label{band}
%\end{figure*}

\subsection{\label{calculation}Numerical Calculations}
The self-consistent equations derived above cannot generally be solved analytically, but the solution can be obtained through numerical iteration \cite{Noro2010-cm}. 
We discretize the wavenumber as follows: 
\begin{align}
dk_j &=  k_{\text{c}}\frac{j}{\sum_{j=1}^{j_{\text{max}}}j}\label{numerical},
\quad
k_j = \frac{1}{2}dk_j + \sum_{j'=1}^{j-1}dk_{j'},
\end{align}
\if0
\begin{align}
k_j &=& \frac{1}{2}dk_j + \sum_{j'=1}^{j-1}dk_{j'}
\end{align}
\fi
where $j=1,2,\dots,j_{\text{max}}$, and $k_{\text{c}}$ represents the cutoff wavenumber. Here, we set $j_{\text{max}}=1000$.
This discretization ensures a higher density of points near the Dirac point, where singular behaviors are expected, allowing for a more precise analysis of the characteristic features in this region.

\section{\label{linear2}Calculated results from self-consistent Born approximation}
The DOS and electrical conductivity are obtained using SCBA.
It is important to note that the results do not explicitly depend on $q_0$ because both the DOS and conductivity are functions of $\mu/(q_0 \hbar v)$, and are normalized by $q_0^2/\hbar v$ and $e^2 q_0/\hbar$, respectively.

\subsection{Dependence on curvature}\label{sec_c}
%------------------------------------------------------------%
\begin{figure*}
\includegraphics[width=16cm]{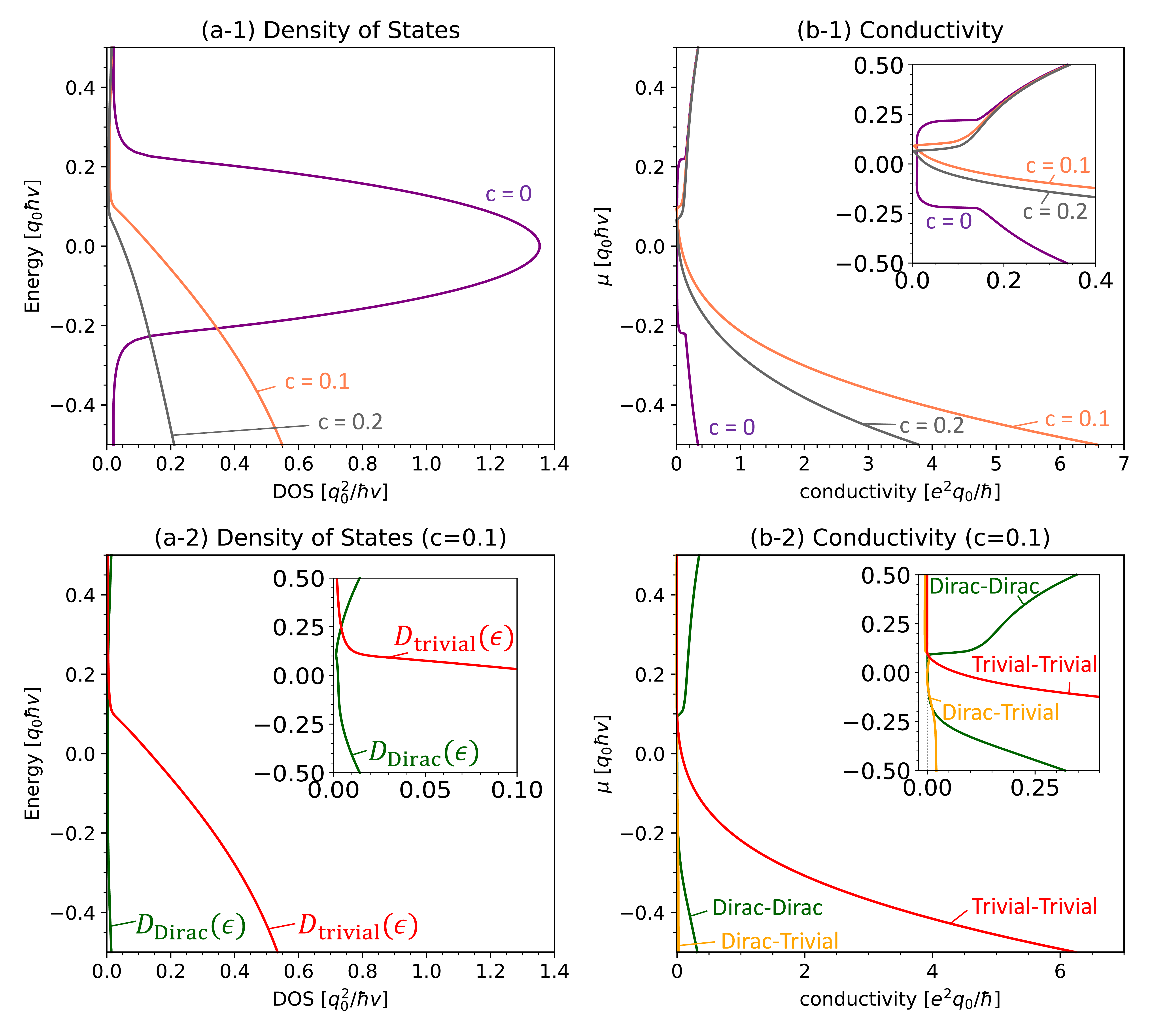}
\caption{(Color online) Quantum transport for $W=2$, $k_{\text{c}}=3q_0$ and $T=0$ derived by the SCBA. 
	(a-1) DOS for $c=0$ (purple line), $c=0.1$ (orange line), and $c=0.2$ (gray line). 
	(a-2) DOS from the Dirac cone (green line) and the trivial band (red line) for $c=0.1$.
	(b-1) Conductivity for $c=0$ (purple line), $c=0.1$ (orange line), and $c=0.2$ (gray line). 
	(b-2) Conductivity from the intraband contribution of the Dirac cone (green line), from the interband contribution between the Dirac cone and the trivial band (yellow line), and the intraband contribution of the trivial band (red line) for $c=0.1$.}
\label{fig_c}
\end{figure*}
%------------------------------------------------------------%
Figure~\ref{fig_c} shows the DOS and electrical conductivity for $W=2$, $k_c=3q_0$, and $T=0$.
For $c=0$, the large DOS peak near the zero energy is the contribution from the completely flat trivial band \cite{Kikuchi2022}. 
On the other hand, for $c>0$, the DOS increases as $\sqrt{-\epsilon}$, dominantly from the trivial band, verified in Fig.~\ref{fig_c}(a-2). 
Furthermore, the DOS is finite even for a small positive value of $\mu$, which is a tail of the trivial band broadened by impurities. 
As $c$ increases, the tail is suppressed. 

Figure~\ref{fig_c}(b-1) shows the electrical conductivity. 
For $c=0$, since the electronic states in $|\mu|<0.25 q_0 \hbar v$ are dominated by the flat trivial band with vanishing velocity, the conductivity is suppressed \cite{Kikuchi2022}. 
In the case of $c>0$ and $\mu>0$, the conductivity shows a similar suppression at $\mu \simeq 0.1 q_0 \hbar v$, as in the case of a flat trivial band ($c=0$). 
Notably, the intraband term derived from the Dirac band is still significantly suppressed in the low-energy region $-0.25 q_0 \hbar v < \mu < 0.1 q_0 \hbar v$.
In contrast, in $\mu<0$, the conductivity increases as $|\mu|$ increases. 
The main contribution comes from the intraband term Eq.~(\ref{intra}) of the trivial band, which has a finite velocity $-2c'k$, as shown in Fig.~\ref{fig_c}(b-2).

\subsection{Dependence on impurity scattering}\label{sec_W}
%------------------------------------------------------------%
\begin{figure*}
\includegraphics[width=16cm]{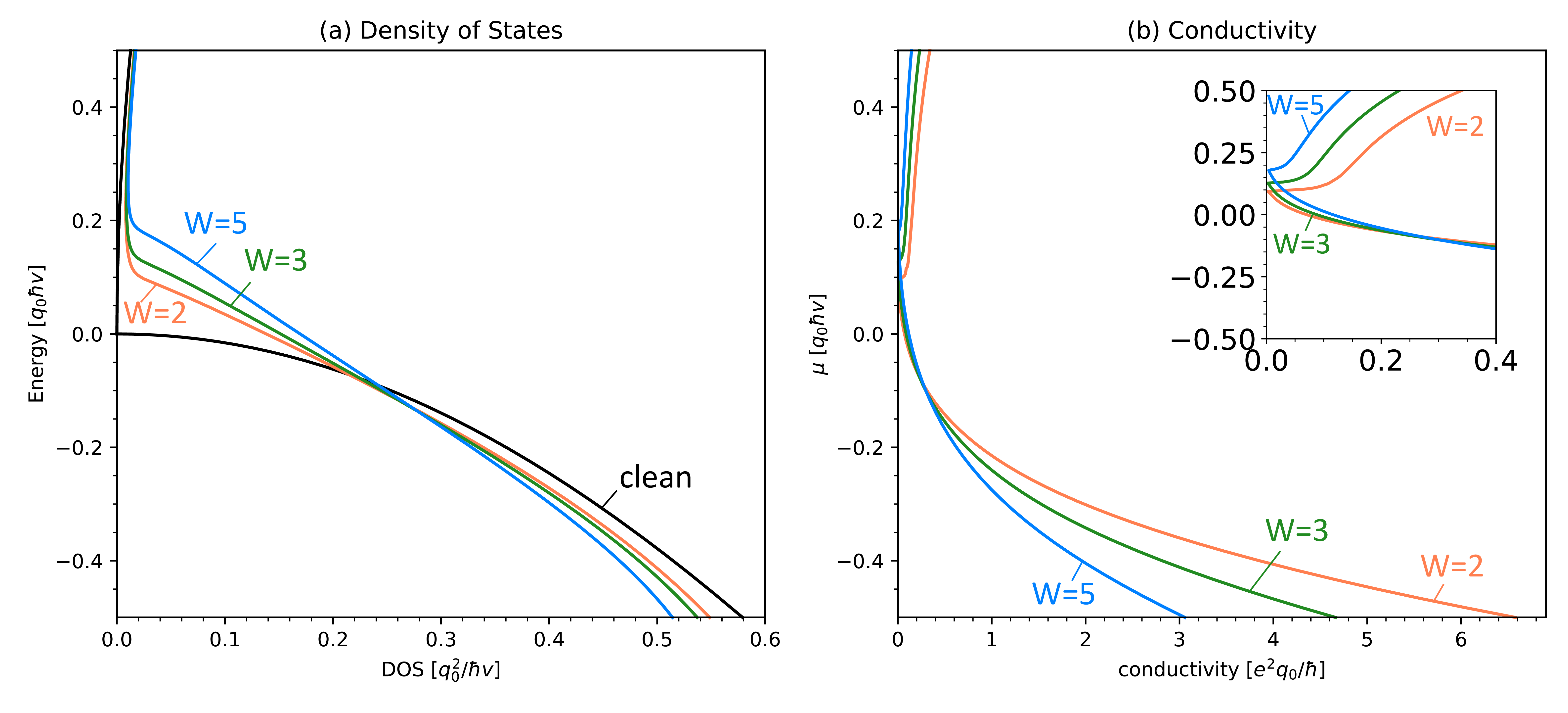}
\caption{(Color online) Quantum transport for $c=0.1$, $k_{\text{c}}=3q_0$ and $T=0$ derived by the SCBA. 
	(a) DOS for the clean system (black line) given by Eq.~(\ref{eq_clean}), $W=2$ (orange line), $W=3$ (green line), and $W=5$ (blue line). 
	(b) The conductivity for $W=2$ (orange line), $W=3$ (green line), and $W=5$ (blue line).}
\label{fig_W}
\end{figure*}
%------------------------------------------------------------%

Figure~\ref{fig_W} shows the DOS and conductivity for $c=0.1$, $k_c=3q_0$, and $T=0$.
As the impurity strength $W$ increases, the DOS increases in $0 < \epsilon < 0.2 q_0 \hbar v$ due to the broadening of the trivial band.
In the high--$|\mu|$ region, the conductivity is reduced by the impurity, as in conventional metals.
We find a region, $\mu \sim 0 - 0.1 q_0 \hbar v$, in which conductivity increases with the strength of the disorder $W$.
This corresponds to the DOS enhancement. 

\subsection{Electrical conductivity in the crossing point between Dirac cone and trivial band}
%------------------------------------------------------------%
\begin{figure}
\includegraphics[width=5cm]{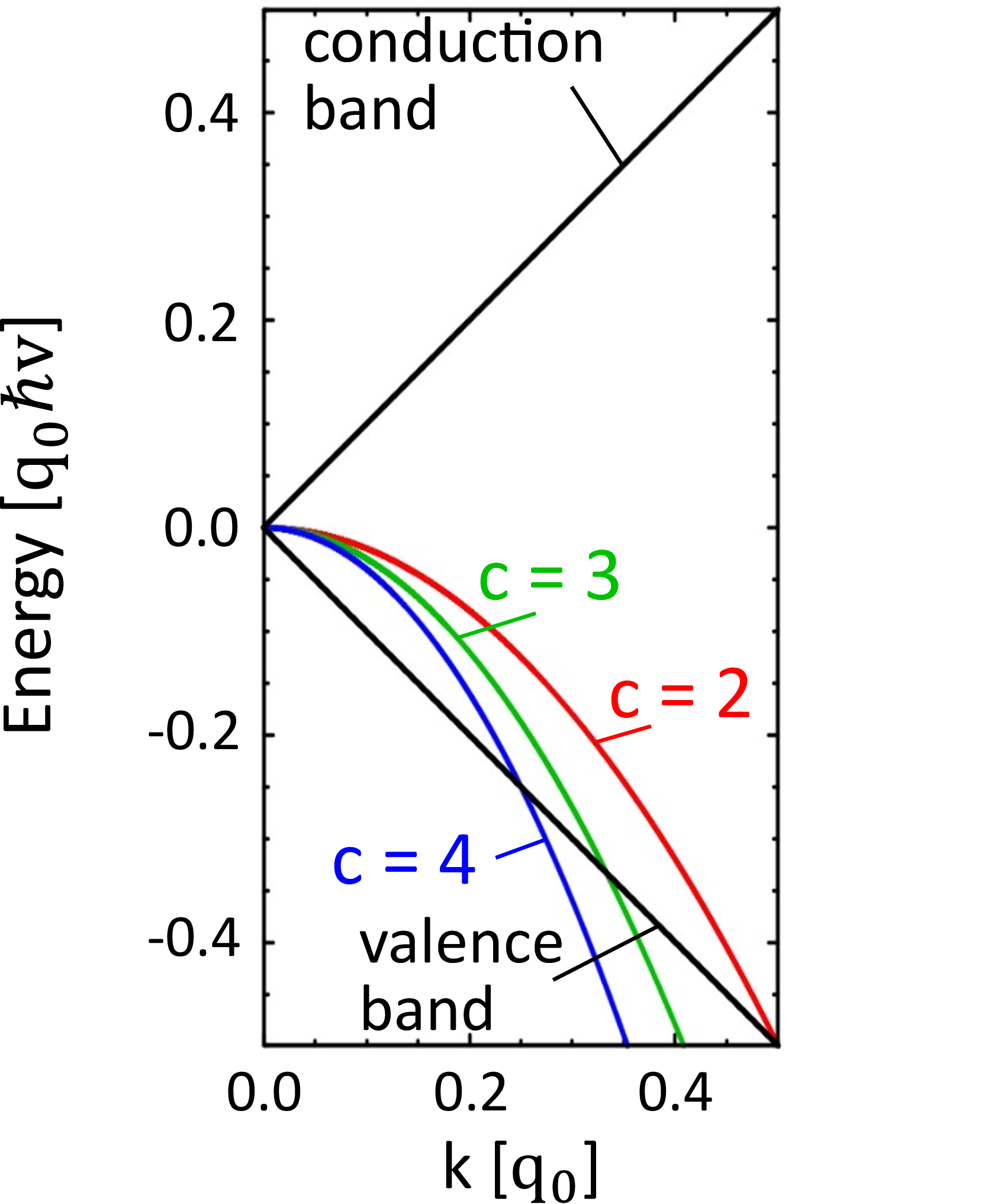}
\caption{(Color online) The energy band of the spin-1 chiral fermion system for $c=2$ (red line), $c=3$ (light green line), and $c=4$ (blue line).}
\label{fig_band2} 
\end{figure}
%------------------------------------------------------------%
%------------------------------------------------------------%
\begin{figure*}
\includegraphics[width=16cm]{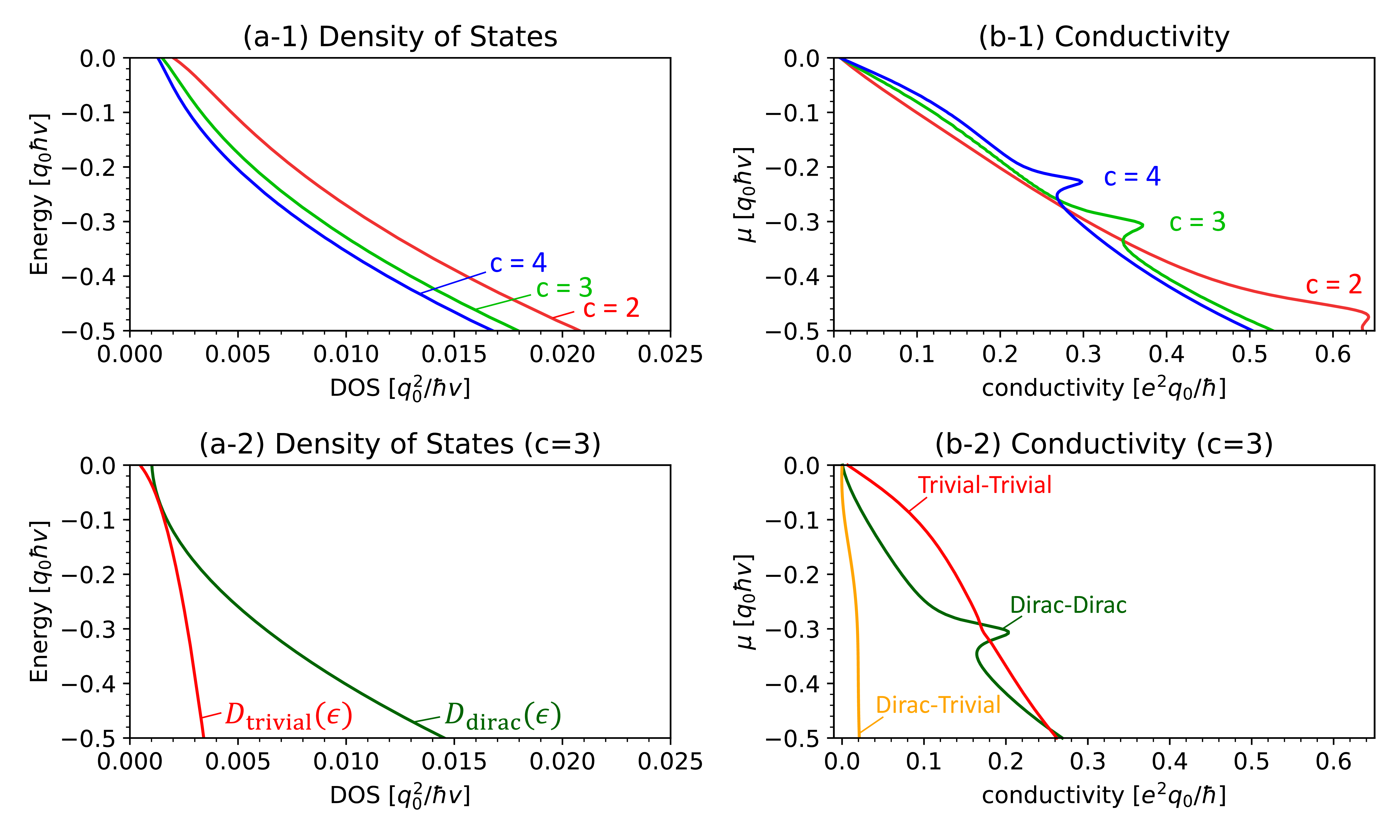}
\caption{(Color online) Quantum transport for $W=2$, $k_{\text{c}}=3q_0$ and $T=0$ derived by the SCBA. 
	(a-1) DOS for $c=2$ (red line), $c=3$ (light green line), and $c=4$ (blue line). 
	(a-2) DOS from the Dirac cone (green line) and the trivial band (red line) for $c=3$.
	(b-1) The conductivity for $c=2$ (red line), $c=3$ (light green line), and $c=4$ (blue line).  
	(b-2) The conductivity from the intraband contribution of the Dirac cone (green line), from the interband contribution between the Dirac cone and the trivial band (yellow line), and from the intraband contribution of the trivial band (red line) for $c=3$.}
\label{fig_c2} 
\end{figure*}
%------------------------------------------------------------%
%------------------------------------------------------------%
\begin{figure}
\includegraphics[width=8cm]{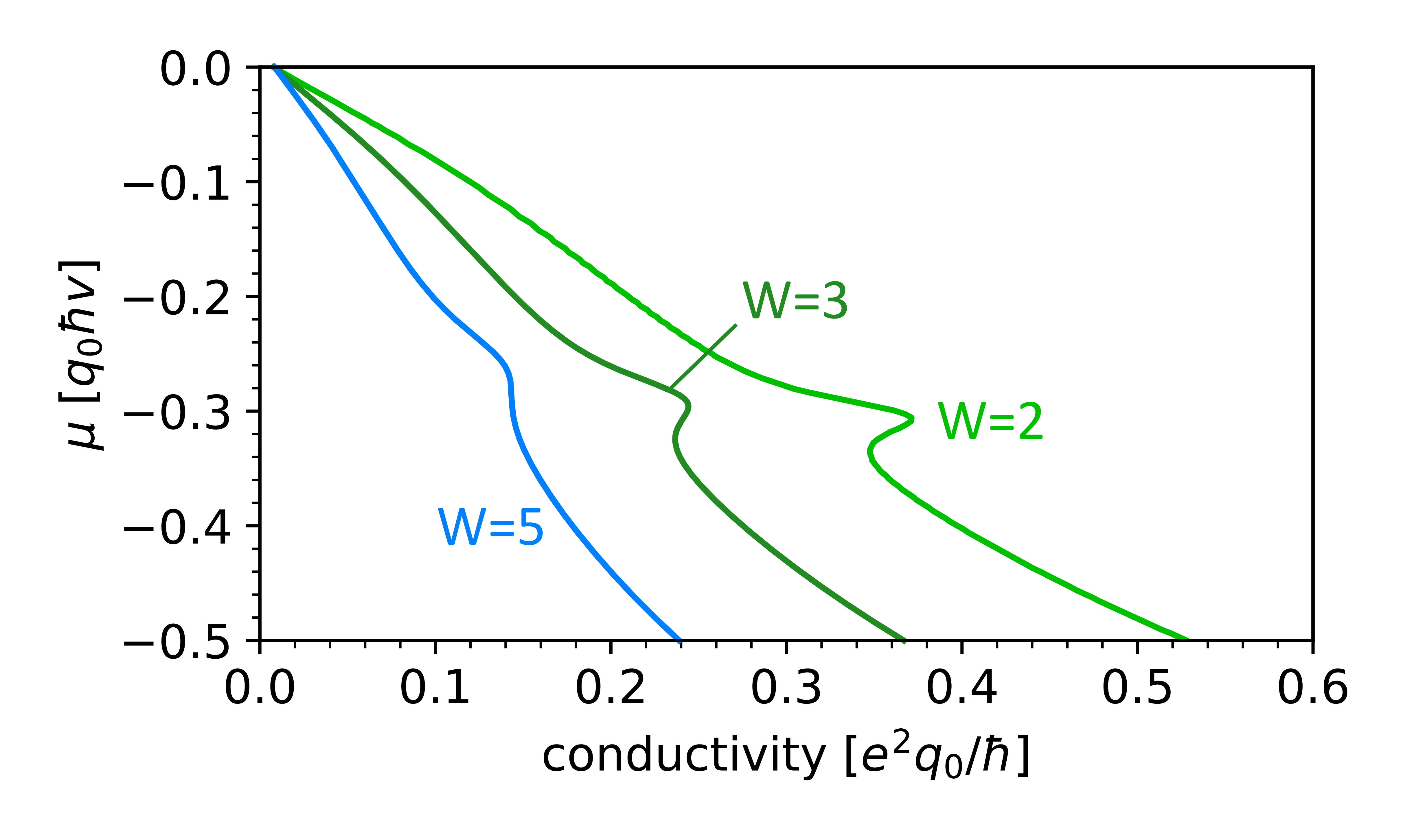}
\caption{(Color online) Conductivity for $W=2$ (light green line), $W=3$ (green line), and $W=5$ (blue line). They are derived by the SCBA for $c=3$, $k_{\text{c}}=3q_0$ and $T=0$.}
\label{fig_W2} 
\end{figure}
%------------------------------------------------------------%
%------------------------------------------------------------%
\begin{figure*}
\includegraphics[width=16cm]{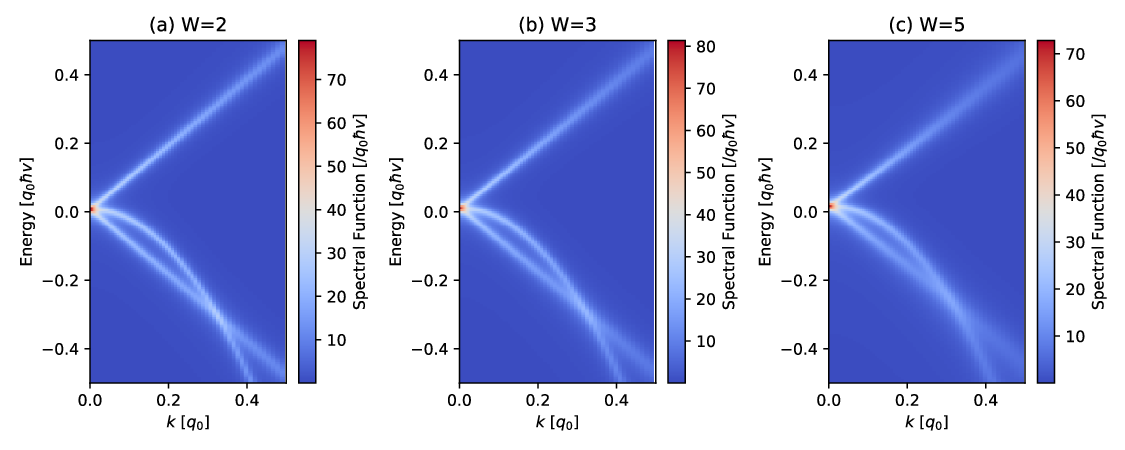}
\caption{(Color online) Spectral function ($c=3$ and $k_{\text{c}}=3q_0$) derived by the SCBA for (a) $W=2$ , (b) $W=3$ and (c) $W=5$.}
\label{fig_spectral}
\end{figure*}
%------------------------------------------------------------%
Next, we discuss the DOS and conductivity in the band-crossing point between Dirac and trivial bands, with fixed parameters $W=2$, $k_c=3q_0$, and $T=0$.  
The Dirac and trivial bands cross at $\mu =-1/c [q_0 \hbar v]$, as shown in Fig.~\ref{fig_band2}. 
The results for the density of states (DOS) and electrical conductivity in $\mu<0$ are shown in Fig.~\ref{fig_c2}.

Figure~\ref{fig_c2}(a-1) shows that the DOS increases as $\epsilon^2$, which means that the Dirac band dominates this energy region, as verified in Fig.~\ref{fig_c2}(a-2).
Figure~\ref{fig_c2}(b-1) shows the electrical conductivity. 
Although the DOS monotonically increases with the energy, the conductivity exhibits a peak near the band-crossing point ($\epsilon=-1/c$[$q_0\hbar v$]). 
By separating the conductivity into intraband and interband contributions for $c=3$ as shown in Fig.~\ref{fig_c2}(b-2), it becomes clear that this peak is primarily caused by the intraband effect of the Dirac cone.

Furthermore, as shown in Fig.~\ref{fig_W2}, as the impurity parameter $W$ increases, the peak of the conductivity shifts away from $\epsilon=-1/c$[$q_0\hbar v$]. 
The reason for this shift is made clear by examining the spectral function $D_{\bm{k}}(\epsilon)$, shown in Fig.~\ref{fig_spectral}: 
\begin{align} 
D_{\bm{k}}(\epsilon) = -\frac{1}{\pi}\Im\Tr\hat{G}(\bm{k},\epsilon+i0). 
\end{align} 
This figure shows that the effective band-crossing point ($\epsilon \sim -0.3 q_0 \hbar v$) shifts upward with increasing $W$ due to the increase in the real part of the self-energy, leading to the conductivity peak that depends on $W$.

\subsection{Temperature dependence}
%------------------------------------------------------------%
\begin{figure*}
\includegraphics[width=18cm]{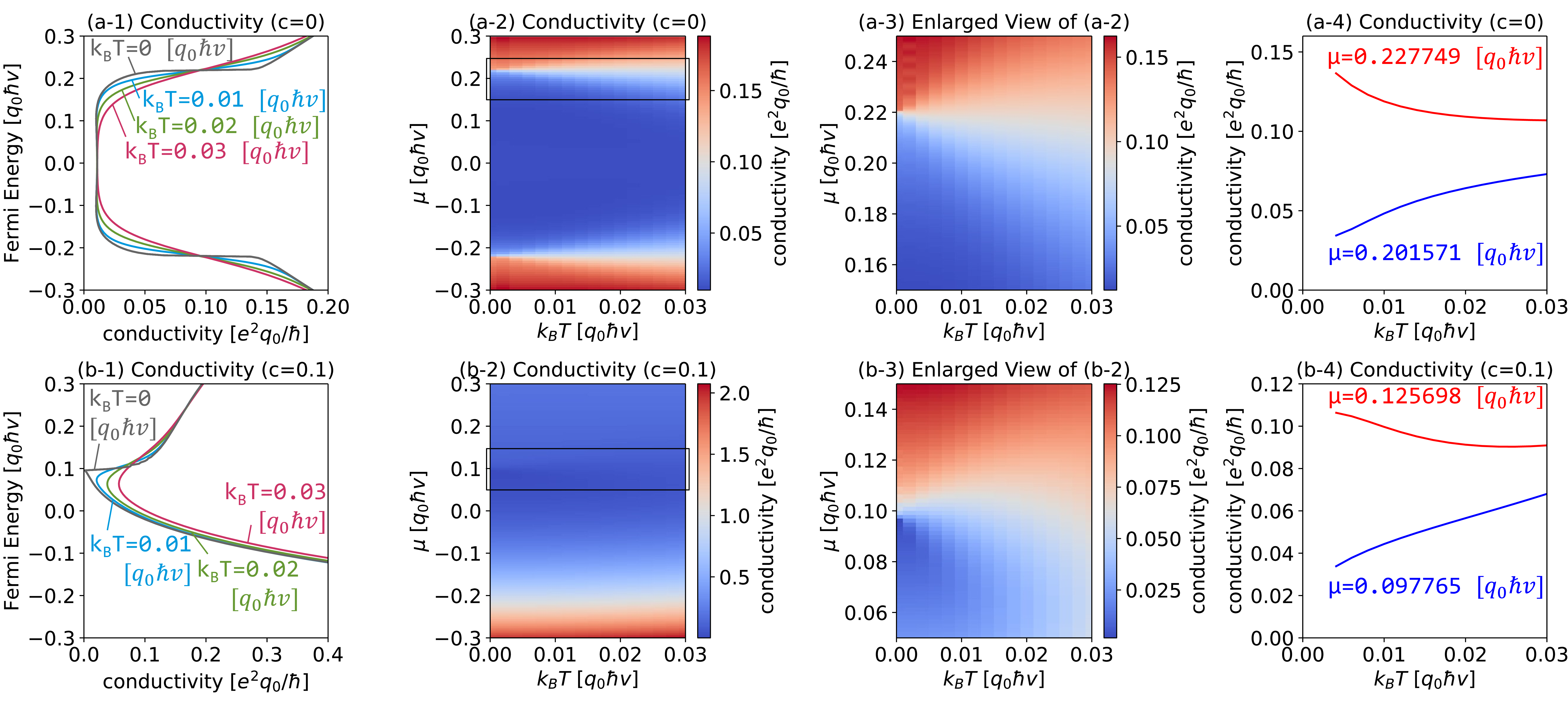}
\caption{
(Color online) Conductivity ($W=2$, $k_{\text{c}}=3q_0$) obtained from the SCBA for (a-1) and (a-2) with $c=0$, and (b-1) and (b-2) with $c=0.1$.
(a-3) is the enlarged view of (a-2), and (b-3) is the enlarged view of (b-2).
(a-4) Conductivity ($c=0$) for $\mu=0.201571~q_0\hbar v$ (blue line) and for $\mu=0.227749~q_0\hbar v$ (red line).
(b-4) Conductivity ($c=0.1$) for $\mu=0.097765~q_0\hbar v$ (blue line) and for $\mu=0.125698~q_0\hbar v$ (red line).}
\label{fig-T}
\end{figure*}
%------------------------------------------------------------%
%------------------------------------------------------------%
\begin{figure*}
\includegraphics[width=16cm]{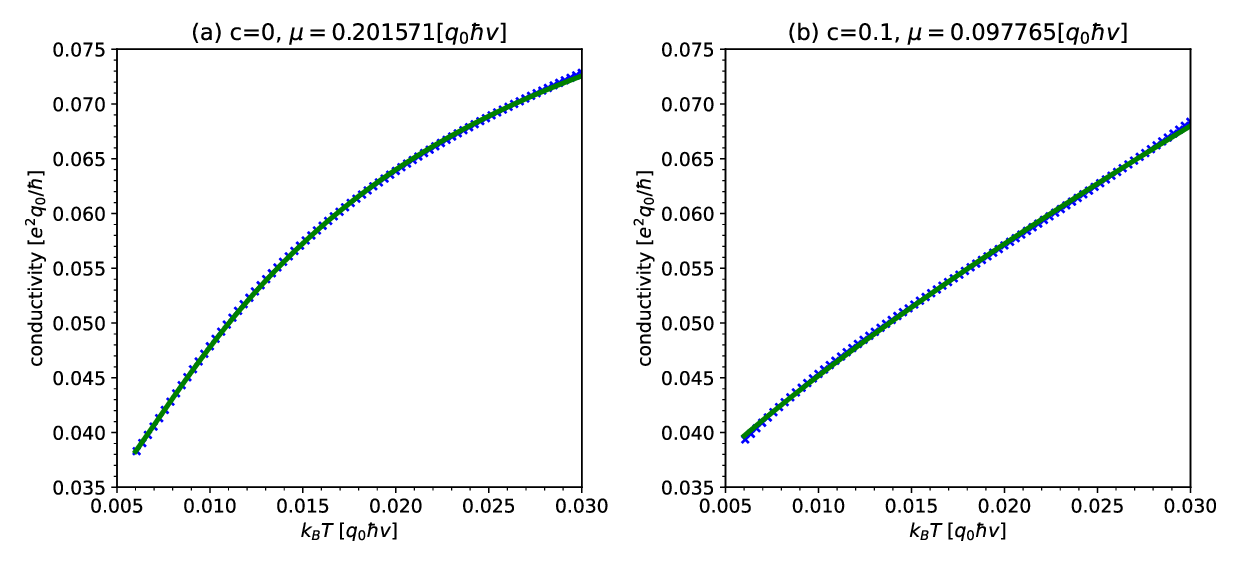}
\caption{(Color online) Fitting of the conductivity in Figs.~\ref{fig-T} (a-4) and \ref{fig-T} (b-4).
(a) The conductivity for $c=0$ and $\mu=0.201571 q_0\hbar v$ as shown in Fig.~\ref{fig-T}(a-4). 
The green solid line shows the fitting function of Eqs.~(\ref{eq-fitting})--(\ref{eq-fittingC}).
(b) The conductivity for $c=0.1$ and $\mu=0.097765 q_0\hbar v$ as shown in Fig.~\ref{fig-T}(b-4). 
The green solid line shows the fitting function of Eqs.~(\ref{eq-fitting2})--(\ref{eq-fitting2C}).
}
\label{fig-fitting} 
\end{figure*}
%------------------------------------------------------------%
%------------------------------------------------------------%
\begin{figure*}
\includegraphics[width=16cm]{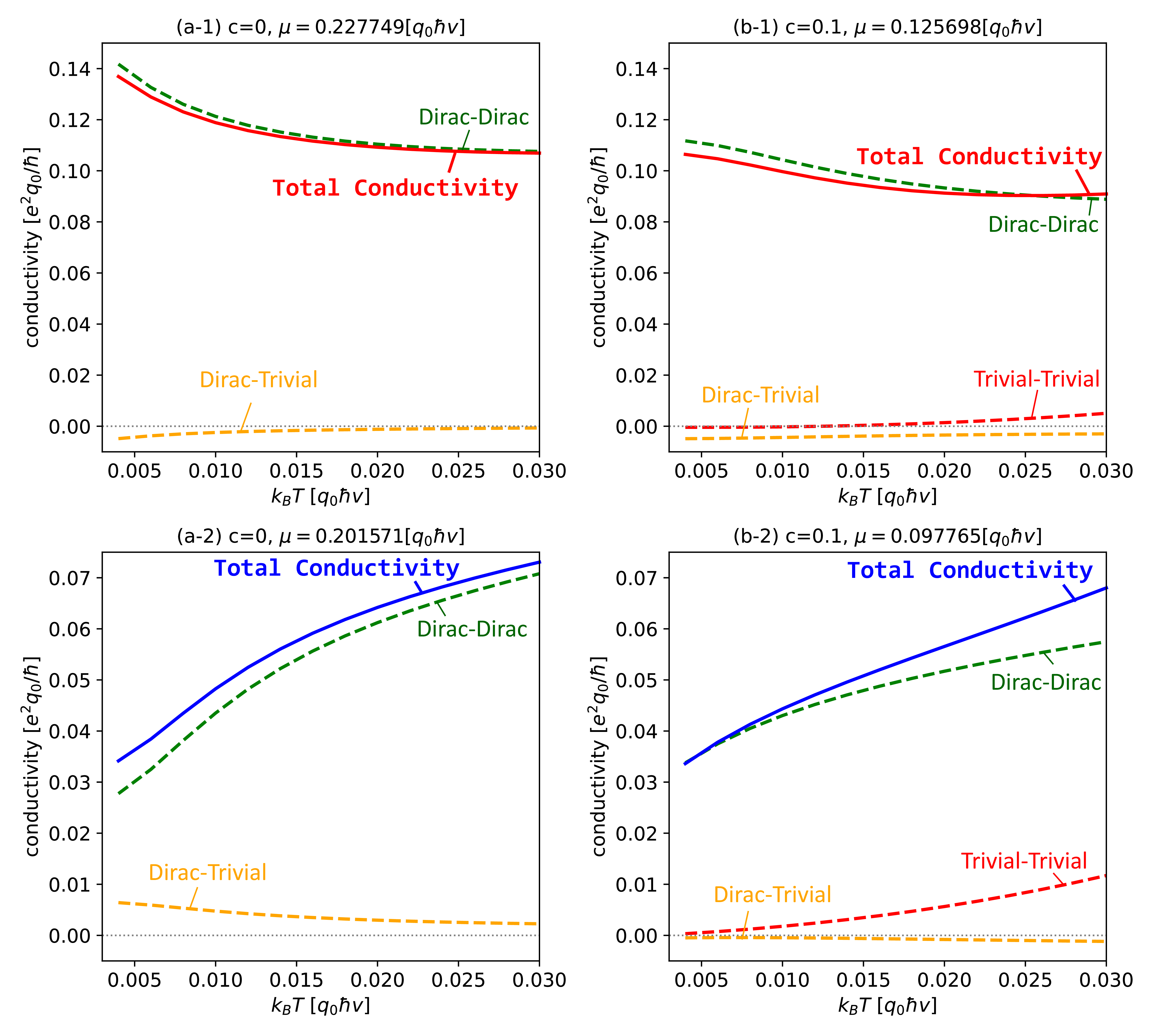}
\caption{
(Color online) Band decomposition of the conductivity in Figs.~\ref{fig-T}~(a-4) and \ref{fig-T}~(b-4) into the intraband contribution of the Dirac cone (green dashed line), interband contribution between the Dirac cone and the trivial band (yellow dashed line), and intraband contribution of the trivial band (red dashed line). 
The total conductivity is represented by a solid line, as shown in Figs.~\ref{fig-T} (a-4) and \ref{fig-T} (b-4). The parameters are as follows: (a-1) $c=0$, $\mu=0.227749 q_0 \hbar v$; (a-2) $c=0$, $\mu=0.201571 q_0 \hbar v$; (b-1) $c=0.1$, $\mu=0.125698 q_0 \hbar v$; (b-2) $c=0.1$, $\mu=0.097765 q_0 \hbar v$.}
\label{fig-T2}
\end{figure*}
%------------------------------------------------------------%

Figure~\ref{fig-T} shows the temperature dependence of the conductivity with fixed parameters $k_c=3q_0$ and $W=2$. 
In this analysis, only impurity scattering is considered, and the discussion applies to the low-temperature regime.
The strong dependence of the conductivity on the chemical potential, which is a characteristic of spin-1 fermion systems, leads to a pronounced temperature dependence of the conductivity, as discussed below.

Figure~\ref{fig-T}(a-1) and (a-2) show the dependence of the conductivity on the chemical potential and temperature for $c=0$. 
It is clear that the temperature dependence varies with the chemical potential. 
the chemical potential range where the Dirac cone dominates at $T=0$ ($0.22~q_0 \hbar v \lessapprox |\mu| \lessapprox 0.3~q_0 \hbar v$), the conductivity decreases as the temperature increases. 
Conversely, in the chemical potential range where the trivial band dominates at $T=0$ ($-0.22~q_0 \hbar v \lessapprox \mu \lessapprox 0.22~q_0 \hbar v$), the conductivity increases with increasing temperature. 
This is more clearly shown in Fig.~\ref{fig-T}(a-3), where a zoomed-in view highlights the behavior. 
As seen in Fig.~\ref{fig-T}(a-4), even with a fixed value of the chemical potential, this difference becomes apparent.

The result for $\mu=0.201571 q_0 \hbar v$ (blue line) in Fig.~\ref{fig-T}(a-4) is fitted using the function:
\begin{align}
\sigma(k_{\text{B}} T)=\alpha e^{-\frac{\beta}{k_{\text{B}} T}}+\gamma, \label{eq-fitting}
\end{align}
where $\alpha$, $\beta$, and $\gamma$ are fitting parameters. 
The obtained values are:
\begin{align}
\alpha &= 0.06489 \pm 0.00008 \, e^2 q_0 /\hbar,\label{eq-fittingA}\\
\beta &= 0.01435 \pm 0.00008 \, \hbar v q_0,\label{eq-fittingB}\\
\gamma &= 0.03234 \pm 0.00012 \, e^2 q_0 /\hbar.\label{eq-fittingC} 
\end{align}
The errors represent the asymptotic standard error. 
The fitted results are shown in Fig.~\ref{fig-fitting}(a). 
The conductivity of a semiconductor with an energy gap $E_g$ is typically proportional to $e^{-\frac{E_g}{2 k_{\text{B}} T}}$. Comparing this with Eqs.~(\ref{eq-fitting}) and (\ref{eq-fittingB}), we find that the conductivity at $\mu \sim 0.201571~q_0\hbar v$ behaves similarly to that of a semiconductor with $\frac{E_g}{2} \sim 0.01435~q_0\hbar v$. 
This value roughly corresponds to the difference between $\mu \sim 0.201571~q_0\hbar v$ and the onset of conductivity, $\mu \sim 0.22~q_0\hbar v$.

Figure~\ref{fig-T}(b-1) and (a-2) show the dependence of the conductivity on the chemical potential and temperature for $c=0.1$. 
Even when the trivial band has curvature, the temperature dependence still varies with the chemical potential. 
This is more clearly seen in Fig.~\ref{fig-T}(b-3), where the zoomed-in view makes this behavior evident. 
In the chemical potential range where the Dirac cone dominates at $T=0$ ($0.105~q_0 \hbar v \lessapprox \mu \lessapprox 0.30~q_0 \hbar v$), the conductivity decreases as the temperature increases. 
On the other hand, in the chemical potential range where the trivial band dominates at $T=0$ ($-0.30~q_0 \hbar v \lessapprox \mu \lessapprox 0.105~q_0 \hbar v$), the conductivity increases with increasing temperature. 
As shown in Fig.~\ref{fig-T}(b-4), this difference is apparent even with a fixed value of the chemical potential.
The conductivity for $\mu=~0.097765 q_0 \hbar v$ shown by the blue line in Fig.~\ref{fig-T}(b-4) is fitted using the function:
\begin{align}
\sigma(k_{\text{B}}T)=\alpha \qty( \frac{k_{\mathrm{B}} T}{\hbar v q_0})^{\beta}+\gamma, 
\label{eq-fitting2}
\end{align}
where $\alpha$, $\beta$, and $\gamma$ are fitting parameters. The obtained values are:
\begin{align}
\alpha &= 0.62272 \pm 0.02013 \, e^2 q_0/\hbar,
\label{eq-fitting2A}\\
\beta &= 0.78722 \pm 0.01142,\\
\gamma &= 0.02860 \pm 0.00034 \, e^2 q_0/\hbar.
\label{eq-fitting2C} 
\end{align}
The errors represent the asymptotic standard error. 
The fitted results are shown in Fig.~\ref{fig-fitting}(b). 
While the conductivity increases exponentially with temperature for $c=0$, it increases algebraically for $c=0.1$. 
This is consistent with the fact that the conductivity at $c=0.1$ does not exhibit a gap structure but rather is similar to that of a zero-gap semiconductor.

Thus, a qualitative contrast in temperature dependence is observed between the case where the trivial band has no curvature ($c=0$) and the case where it has curvature ($c=0.1$). 
This difference originates from the intraband effect of the trivial band. 
The conductivity shown in Figs.~\ref{fig-T}(a-4) and \ref{fig-T}(b-4) is decomposed into the intraband and interband contributions, as illustrated in Fig.~\ref{fig-T2}. 
For the flat-band case ($c=0$), the intraband term vanishes. 
By contrast, when the trivial band has curvature $(c = 0.1)$, once the temperature increases and the thermal excitations into the trivial band come into play, the intraband term becomes large, as shown in Figs.~\ref{fig-T2}(b-1) and \ref{fig-T2}(b-2). 
These results suggest that the curvature in the trivial band significantly influences the temperature dependence of conductivity.

\section{BOLTZMANN TRANSPORT THEORY}\label{Boltzmann}
In the Boltzmann equation approach, the qualitative behavior of the intraband effect can be derived more easily than in the SCBA. Furthermore, it is possible to express these effects in an analytical form.
In this section, we analytically derive the conductivity at absolute zero temperature using Boltzmann theory. We then compare these results with those obtained from the SCBA.
Note that the calculations in this section are performed under the assumption that $c>0$ and $\epsilon \neq 0$.

%------------------------------------------------------------%
\begin{figure*}
\includegraphics[width=18cm]{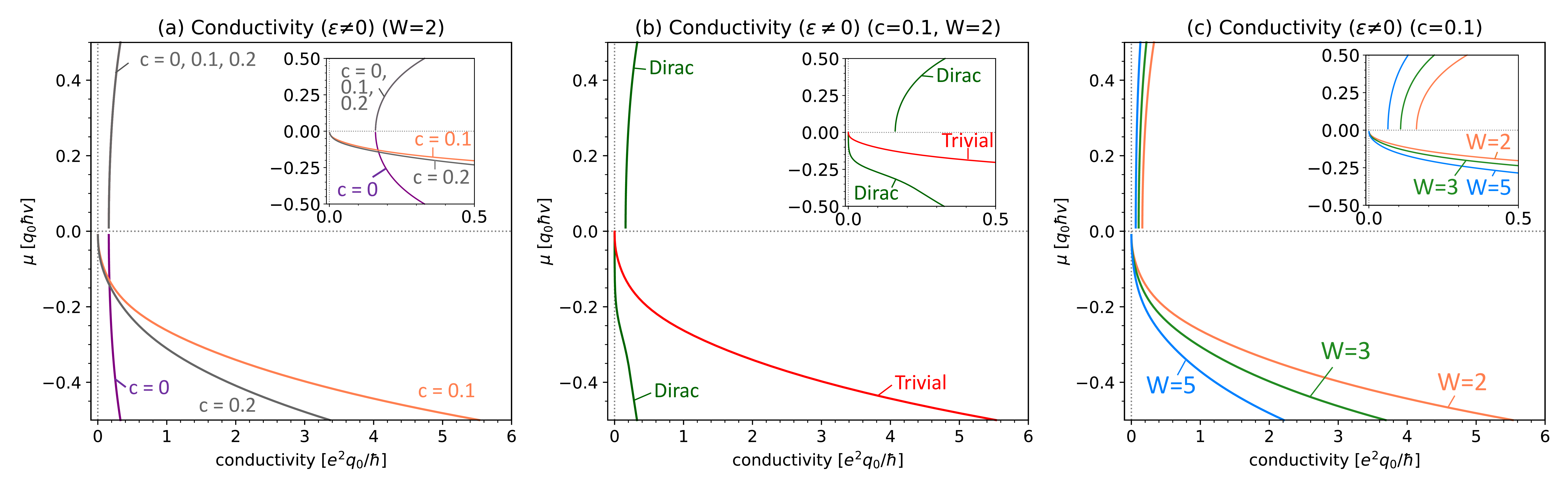}
\caption{(Color online) Conductivity for $T=0$ derived by the Boltzmann equation. 
	(a) Conductivity ($W=2$) for $c=0$ \cite{Kikuchi2022} (purple line), $c=0.1$ (orange line), and $c=0.2$ (gray line). 
	(b) Conductivity from the Dirac cone (green line) and the trivial band (red line) for $c=0.1$ and $W=2$.
	(c) Conductivity ($c=0.1$) for $W=2$ (orange line), $W=3$ (green line), and $W=5$ (blue line).}
\label{fig_boltzmann}
\end{figure*}
%------------------------------------------------------------%

\subsection{Formulation}
The conductivity is given by 
\begin{align}
	\sigma_{\text{B}}(\mu) = 
 \sum_\lambda
 e^2 \int d\epsilon 
 \left(-\frac{\partial f}{\partial \epsilon}\right) 
 v_x^\lambda(\epsilon)^2 
 \tau_{\text{tr}}^\lambda(\epsilon) D_0^\lambda(\epsilon),
	\label{conductivity_B}
\end{align}
where $D_0^\lambda(\epsilon)$ is the DOS per unit volume in the clean limit, $\tau_{\mathrm{tr}}^\lambda(\epsilon)$ is the transport relaxation time 
\begin{align}
v_x^\lambda(\epsilon)=\frac{1}{\hbar} \frac{\partial \epsilon_{\lambda,\bm{k}}}{\partial k_x},
\end{align}
 for each band $\lambda= \mathrm{c}, \mathrm{t}, \mathrm{v}$.
$D_0^\lambda(\epsilon)$ is given by
\begin{align}
  &D^{\mathrm c}_{0}(\epsilon)=\frac{\epsilon^2}{2\pi^2(\hbar v)^3}\theta(\epsilon),
  \\
  &D^{\mathrm t}_{0}(\epsilon)=\frac{q_0}{4\pi^2 c \hbar v}\sqrt{-\frac{q_0\epsilon}{c\hbar v}}\theta(-c\epsilon), 
  \\
  &D^{\mathrm v}_{0}(\epsilon)=\frac{\epsilon^2}{2\pi^2(\hbar v)^3}\theta(-\epsilon).
\end{align}
for each energy band.

The transport relaxation time is calculated by
\begin{align}
\frac{1}{\tau_{\rm{tr}}^\lambda(\epsilon)} = \sum_{\lambda'}\int\frac{d\bm{k'}}{(2\pi)^3}(1-\cos\theta_{\bm{k'}\bm{k}})W_{\lambda'\bm{k'}, \lambda\bm{k}}. \label{tau}
\end{align}
where $\theta_{\bm{kk'}}$ is the angle between $\bm{k}$ and $\bm{k'}$, and the scattering probability $W_{\lambda'\bm{k'},\lambda \bm{k}}$ is given by Fermi's golden rule:
\begin{align}
W_{\lambda'\bm{k'}, \lambda\bm{k}} = \frac{2\pi}{\hbar}n_{\text{i}}|
\bra{\lambda',\bm{k'}}U\ket{\lambda,\bm{k}}|^2\delta(\epsilon_{\lambda',\bm{k'}}-\epsilon_{\lambda,\bm{k}}),\label{golden_rule}
\end{align}
with
\begin{align}
 \bra{\lambda',\bm{k'}}U\ket{\lambda,\bm{k}}
 = \int d\vb*{r} e^{-i (\vb*{k}-\vb*{k}') \cdot \vb*{r}}
 U(\vb*{r}) \vb*{\xi}_{\lambda', \vb*{k'}}^\dag \vb*{\xi}_{\lambda, \vb*{k}}.
\end{align}
The eigenvector $\boldsymbol{\xi}_{\lambda, \boldsymbol{k}}$ for the band $\lambda$ is given by Eqs.~(\ref{eigenstates1})--(\ref{eigenstates2}).

For the conduction band, the zero-temperature conductivity for $\mu>0$ is 
\begin{gather}
\sigma_{\mathrm{B},\text{c}}(\mu)
= \frac{512}{3\pi W}\dfrac{e^2 q_0}{ \hbar}A\left(\dfrac{\mu}{\hbar vq_0}\right),\\
A(x)=\dfrac{x^8}{32x^4-16x^2+3-(8x^2+3)\exp(-8x^2)}.
\end{gather}
For the trivial band, the zero-temperature conductivity for $\mu<0$ is
\begin{align}
&\sigma_{\mathrm{B},\text{t}}(\mu)=\frac{512}{3\pi W}\dfrac{e^2 q_0}{ \hbar}B\left(\dfrac{-\mu}{\hbar vq_0}\right),
\end{align}
where 
\begin{align}
&B(x)=\frac{x^6}{B_0(x)},\\
&B_0(x)=c x(8 x^2 - 8 c x + 3 c^2)
\nonumber\\
&-x \left(64 x^3 + 40 c x^2 + 16 c^2 x + 3 c^3 \right)
\exp\left(-8\frac{x}{c}\right)
\nonumber\\
&+c^2 \left(8 x^2 - 3 c \sqrt{\frac{x}{c}}\right)
\exp\left(-2x^2 + 4 x\sqrt{\frac{x}{c}}- 2\frac{x}{c}\right)
\nonumber\\
&+c^2\left( 32 x^3 \sqrt{\frac{x}{c}} +16 x^2 + 3 c \sqrt{\frac{x}{c}} \right) 
\notag\\&\hspace{4em}\times
\exp\left(-2x^2 - 4 x\sqrt{\frac{x}{c}}- 2\frac{x}{c}\right).
\end{align}
For the valence band, the zero-temperature conductivity is 
\begin{align}
\sigma_{\mathrm{B},\text{v}}(\mu)=\frac{512}{3\pi W}\dfrac{e^2 q_0}{ \hbar}C\left(\dfrac{-\mu}{\hbar vq_0}\right),
\end{align}
where
\begin{align}
&C(x)=\frac{x^8}{C_0(x)},\\
&C_0(x)=32x^4 - 16 x^2 + 3 -(8 x^2+ 3)\exp\left(-8 x^2\right)\nonumber\\
&+\left(8 x^2 - 3 c \sqrt{\frac{x}{c}}\right)\exp\left(-2x^2 + 4x \sqrt{\frac{x}{c}} - 2\frac{x}{c}\right)
\notag\\& + 
\left(32 x^3 \sqrt{\frac{x}{c}} + 16 x^2 + 3 c \sqrt{\frac{x}{c}}\right)
\notag\\&\hspace{4em}\times
\exp\left(-2x^2 - 4x \sqrt{\frac{x}{c}} - 2\frac{x}{c}\right).
\end{align}
The conductivity for the spin-1 chiral fermion system derived from the Boltzmann equation is then given by 
\begin{align}
\sigma_\text{B}(\mu)=\sigma_{\mathrm{B},\text{c}}(\mu)\theta(\mu)+(\sigma_{\mathrm{B},\text{t}}(\mu)+\sigma_{\mathrm{B},\text{v}}(\mu))\theta(-\mu).
\end{align}
The results are shown in Fig.~\ref{fig_boltzmann}.
Note that $c=0$ is singular as
\begin{align}
 \lim_{c \to 0} \sigma_{\mathrm{B}, \mathrm{t}}(\mu)
 = \infty.
\end{align}

\subsection{Comparison of Boltzmann equation and SCBA}
\begin{figure*}
\includegraphics[width=16cm]{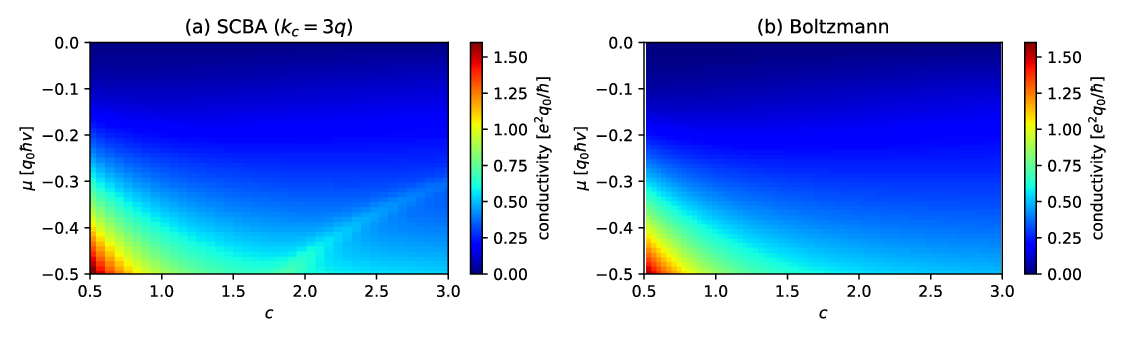}
\caption{(Color online) Conductivity for $T=0$ derived by the SCBA ($k_{\text{c}}=3q_0$) (a) and the Boltzmann equation (b).}
\label{fig-c-e-conductivity2} 
\end{figure*}
%------------------------------------------------------------%

There are two major differences between the conductivity derived from the SCBA and the Boltzmann equation.
First, the Boltzmann conductivity is singular at $\mu=0$, independent of $W$ and $c$. 
For $\mu<0$, the conductivity is dominated by the trivial band, while for $\mu>0$, only the Dirac band contributes to the conductivity. 
The conductivity calculated by SCBA has no singular point and changes abruptly at the finite value of $\mu = \mu_0$, which depends on $W$ and $c$. 
The conductivity is dominated by the intraband contribution of the trivial band for $\mu < \mu_0$ and the Dirac band for $\mu > \mu_0$. 
In the Boltzmann theory, $\mu_0=0$ is realized.
Next, as seen in Fig.~\ref{fig-c-e-conductivity2}, the enhancement in conductivity at the crossing point of the Dirac and trivial bands for $k \ne 0$, which appear for $c > 2$ in Fig.~\ref{fig-c-e-conductivity2}(a), is not reproduced in the Boltzmann theory.
This implies that the higher-order contribution in the SCBA self-energy is responsible for the emergence of the enhancement.
These distinctions suggest that SCBA is indeed effective in evaluating conductivity behavior in the low-energy region and near the band-crossing point.

\section{\label{discussion}Discussion}
%------------------------------------------------------------%
\begin{figure*}
\includegraphics[width=18cm]{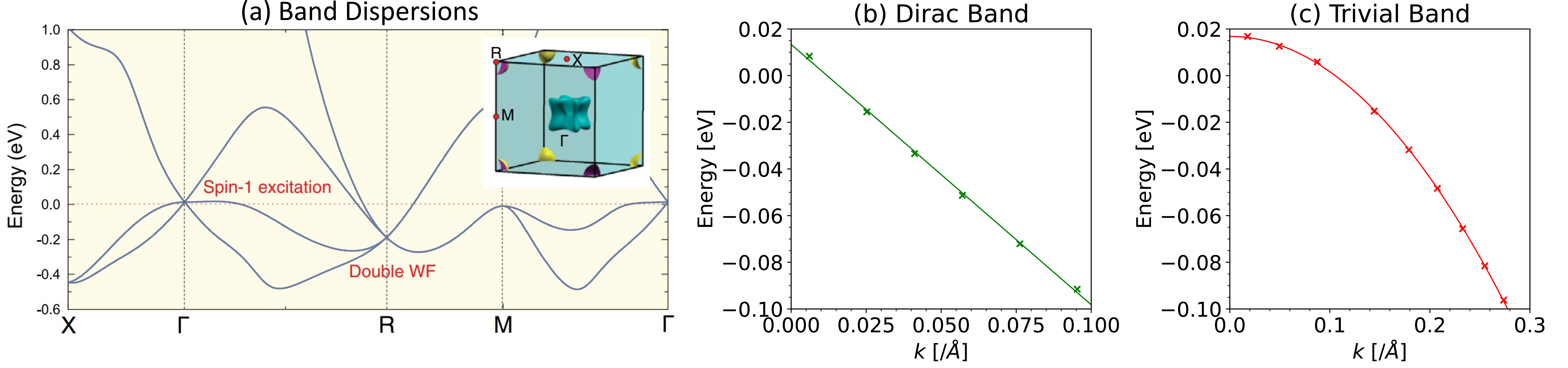}
\caption{(Color online) (a) Energy band of CoSi. The figure is from Tang \textit{et al}., 2017 \cite{Tang2017-kk}. 
(b) and (c) The symbols show the dispersion of (b) the Dirac cone and (c) the trivial band along the $\Gamma X$ cut, and the solid lines show the best fits as given in Eqs.~(\ref{eq-fitting3A})--(\ref{eq-fitting3B}) and Eqs.~(\ref{eq-fitting4A})--(\ref{eq-fitting4B}).
}
\label{parameter}
\end{figure*}
%------------------------------------------------------------%

We have theoretically clarified the energy dependence of the DOS and the conductivity. 
These quantities can be experimentally measured in thin films of spin-1 chiral fermion materials by continuously tuning the energy via gating. 
Alternatively, the chemical potential can be discretely tuned by doping in bulk materials. 
%However, in real materials, in addition to the spin-1 fermion system, there are electronic states characterized by the double Weyl fermion system. To fully understand transport measurements, it is essential to understand the quantum transport phenomena associated with the double Weyl fermion system.

Here, we evaluate the material parameter of CoSi, which hosts a spin-1 fermion state.
The lattice constant is assumed to be $4.438 \text{\AA}$ \cite{Takane2019}.
The Dirac and trivial bands derived from first-principles calculation for CoSi are fitted by $\epsilon_{\text{Dirac}} = -\hbar v k - \epsilon_{\mathrm{F}}$ and $\epsilon_{\text{trivial}} = -c'k^2 - \epsilon_{\mathrm{F}}$ along the $\Gamma X$ cut in the range of $-0.10 ~\text{eV} < \epsilon < 0.02 ~\text{eV}$, as shown in Fig.~\ref{parameter}.
The fitting parameters are determined as
\begin{align}
\hbar v &= 1.1160 \pm 0.0188 ~\text{eV} \cdot \text{\AA}, \label{eq-fitting3A}\\
\epsilon_{\mathrm{F}} &= -0.0134 \pm 0.0011 ~\text{eV}, \label{eq-fitting3B}
\end{align}
and
\begin{align}
 c' &= 1.5137 \pm 0.0060 ~\text{eV} \cdot \text{\AA}^2, \label{eq-fitting4A}\\
\epsilon_{\mathrm{F}} &= -0.0168 \pm 0.0003 ~\text{eV}. \label{eq-fitting4B}
\end{align}
where the error represents the asymptotic standard error. 

From these results, we estimate the Fermi energy as $\epsilon_{\text{F}} \sim -0.01$ ~eV and the Fermi velocity as $v \sim 10^5$ ~m/s. Furthermore, assuming $q_0 \sim 0.01 \text{\AA}^{-1}$ \footnote{$q_0$ is estimated as the Thomas-Fermi wavelength as the dielectric constant $\kappa \sim 10$, effective coupling constant $\alpha \sim 0.01-0.1$, $q_0 = \sqrt{(4\pi e^2/\kappa) D(\epsilon)} \sim 0.01 \text{\AA}^{-1}$ \cite{Kikuchi2023}.}, we estimate $c \sim 0.01$.

\section{\label{conclusion}Conclusion}
In this study, we have developed a quantum transport theory for spin-1 chiral fermion systems with disordered potential, applying the self-consistent Born approximation (SCBA) and current vertex corrections. Specifically, our work extends the previous theoretical framework \cite{Kikuchi2022} by introducing a quadratic dispersion for the trivial band and considering finite-temperature effects, making the model more applicable to realistic conditions.

We discovered that the conductivity is significantly suppressed near the edge of the trivial band. 
As a result, the conductivity exhibits a sharp change at a finite chemical potential $\mu_0$, where the dominant conducting states shift from the trivial band to the Dirac band. 
To properly capture this behavior, it is essential to account for the broadening of the DOS of the trivial band, with SCBA being a convenient approach to incorporate this effect.
These findings highlight the critical role of impurity scattering when analyzing the transport properties of materials hosting spin-1 chiral fermions, particularly in low-temperature regimes.

\begin{acknowledgments}
	This work is supported by JSPS KAKENHI for Grants (Grant Nos.~JP20K03835 and JP24H00853), and it is also financially supported by JST SPRING, Grant Number JPMJSP2125.
    RK would like to take this opportunity to thank the ``THERS Make New Standards Program for the Next Generation Researchers.''
    %RK would like to take this opportunity to thank the ``Nagoya University Interdisciplinary Frontier Fellowship'' supported by Nagoya University and JST, the establishment of university fellowships towards the creation of science technology innovation, Grant Number JPMJFS2120.

\end{acknowledgments}

\appendix

\section{Useful relations}\label{integral}
Isotropic disorder potential is characterized by the moment of the scattering angle as follows: 
\begin{align} 
V_l^2(k,k') = 2\pi\int_{-1}^1 d(\cos\theta_{\bm{kk'}})|u(\bm{k}-\bm{k'})|^2\cos^l\theta_{\bm{kk'}}, \label{V} 
\end{align} 
where $l$ represents the power of  $\cos^l\theta_{\bm{kk'}}$ in the integrand, reflecting the contribution of different angular components to the scattering potential.

Let $\bm{n}_{\perp 1}$, $\bm{n}_{\perp 2}$, and $\bm{n}$ be three mutually perpendicular unit vectors in three dimensions.
Let $\hat{S}_x$, $\hat{S}_y$, and $\hat{S}_z$ denote the $3 \times 3$ spin-1 representation matrices introduced in the main text. The following useful relations can be derived: 
\begin{align}
	&
(\hat{\bm{S}}\cdot\bm{n})^3=(\hat{\bm{S}}\cdot\bm{n}),
\label{UR1}
\\&
(\hat{\bm{S}}\cdot\bm{n})^2\hat{S}_{i}(\hat{\bm{S}}\cdot\bm{n})^2=(\hat{\bm{S}}\cdot\bm{n})\hat{S}_{i}(\hat{\bm{S}}\cdot\bm{n})=n_{i}(\hat{\bm{S}}\cdot\bm{n}),
\\&
(\hat{\bm{S}}\cdot\bm{n}_{\perp 1})^2+(\hat{\bm{S}}\cdot\bm{n}_{\perp 2})^2+(\hat{\bm{S}}\cdot\bm{n})^2=2\hat{S}_0,
\\&
(\hat{\bm{S}}\cdot\bm{n}_{\perp 1})\hat{S}_{i}(\hat{\bm{S}}\cdot\bm{n}_{\perp 1})+(\hat{\bm{S}}\cdot\bm{n}_{\perp 2})\hat{S}_{i}(\hat{\bm{S}}\cdot\bm{n}_{\perp 2})
\nonumber\\&
+(\hat{\bm{S}}\cdot\bm{n})\hat{S}_{i}(\hat{\bm{S}}\cdot\bm{n})=\hat{S}_{i},
\\&
(\hat{\bm{S}}\cdot\bm{n}_{\perp 1})\hat{S}_{i}\hat{S}_{j}(\hat{\bm{S}}\cdot\bm{n}_{\perp 1})+(\hat{\bm{S}}\cdot\bm{n}_{\perp 2})\hat{S}_{i}\hat{S}_{j}(\hat{\bm{S}}\cdot\bm{n}_{\perp 2})
\nonumber\\&
+(\hat{\bm{S}}\cdot\bm{n})\hat{S}_{i}\hat{S}_{j}(\hat{\bm{S}}\cdot\bm{n})=-\hat{S}_{j}\hat{S}_{i}+2\delta_{ij}\hat{S}_0.
\label{UR5}
\end{align}

An arbitrary unit vector $\boldsymbol{n}'$ can be expressed as
\begin{align}
	\bm{n'}
	=\bm{n}_{\perp 1}\sin\theta\cos\phi+\bm{n}_{\perp 2}\sin\theta\sin\phi+\bm{n}\cos\theta,
\end{align}
where $\theta$ represents the angle between $\bm{n}$ and $\bm{n'}$, and $\phi$ denotes the azimuthal angle in the $\bm{n}_{\perp1}$-$\bm{n}_{\perp2}$ plane.

By applying Eqs.~(\ref{UR1})--(\ref{UR5}), the following integrals can be computed as
\begin{align}
	&
\int_0^{2\pi}\int_0^\pi d\theta d\phi|u(\bm{k}-\bm{k'})|^2(\hat{\bm{S}}\cdot\bm{n'})
=(\hat{\bm{S}}\cdot\bm{n})V_1^2(k,k'),\label{int1}
\\&
\int_0^{2\pi}\int_0^\pi d\theta d\phi|u(\bm{k}-\bm{k'})|^2(\hat{\bm{S}}\cdot\bm{n'})^2
\nonumber\\&
=\left(\frac{3}{2}V_2^2(k,k')-\frac{1}{2}V_0^2(k,k')\right)(\hat{\bm{S}}\cdot\bm{n})^2
\nonumber\\&\quad
+\left(V_0^2(k,k')-V_2^2(k,k')\right)\hat{S}_0,
\\&
\int_0^{2\pi}\int_0^\pi d\theta d\phi|u(\bm{k}-\bm{k'})|^2n'_x\hat{S}_0
=V_1^2(k,k')n_x\hat{S}_0,
\\&
\int_0^{2\pi}\int_0^\pi d\theta d\phi|u(\bm{k}-\bm{k'})|^2n'_x(\hat{\bm{S}}\cdot\bm{n'})
\nonumber\\&
=\left(\frac{3}{2}V_2^2(k,k')-\frac{1}{2}V_0^2(k,k')\right)n_x(\hat{\bm{S}}\cdot\bm{n})
\nonumber\\&\quad
+\frac{1}{2}(V_0^2(k,k')-V_2^2(k,k'))\hat{S}_{x},
\\&
\int_0^{2\pi}\int_0^\pi d\theta d\phi|u(\bm{k}-\bm{k'})|^2n'_x(\hat{\bm{S}}\cdot\bm{n'})^2
\nonumber\\&
=\frac{1}{2}(V_1^2(k,k')-V_3^2(k,k'))\hat{S}_{x}(\hat{\bm{S}}\cdot\bm{n})
\nonumber\\&\quad
+\frac{1}{2}(V_1^2(k,k')-V_3^2(k,k'))(\hat{\bm{S}}\cdot\bm{n})\hat{S}_{x}
\nonumber\\&\quad
+(V_1^2(k,k')-V_3^2(k,k'))n_x\hat{S}_0
\nonumber\\&\quad
+\left(\frac{5}{2}V_3^2(k,k')-\frac{3}{2}V_1^2(k,k')\right)n_x(\hat{\bm{S}}\cdot\bm{n})^2.\label{int2}
\end{align}

\section{Detailed calculations}\label{calculations}

Since $(\hat{\bm{S}}\cdot\bm{n})^3=(\hat{\bm{S}}\cdot\bm{n})$ holds for the spin-1 representation matrices, the self-energy is decomposed into three terms as
\begin{align}
\hat{\Sigma}(\bm{k},\epsilon)
 &= \Sigma_1(k,\epsilon)\hat{S}_0
% \notag\\&\quad
 +\Sigma_2(k,\epsilon)(\hat{\bm{S}}\cdot\bm{n})+\Sigma_3(k,\epsilon)(\hat{\bm{S}}\cdot\bm{n})^2.\label{self energy2}
\end{align}
With the above expansion, the Green's function Eq.~(\ref{green function}) is rewritten as
\begin{align}
\hat{G}(\bm{k},\epsilon)
&= \frac{1}{X(k,\epsilon) \hat{S}_0+Y(k,\epsilon) \hat{\bm{S}}\cdot\bm{n}+Z(k,\epsilon) (\hat{\bm{S}}\cdot\bm{n})^2}
\notag\\
&= x(k,\epsilon) \hat{S}_0+y(k,\epsilon) (\hat{\bm{S}}\cdot\bm{n})+z(k,\epsilon) (\hat{\bm{S}}\cdot\bm{n})^2\label{green function2},
\end{align}
where
\begin{align}
X(k,\epsilon)&=\epsilon{+\frac{c\hbar v}{q_0}k^2}-\Sigma_1(k,\epsilon),
\\
Y(k,\epsilon)&=-\hbar vk-\Sigma_2(k,\epsilon),
\\
Z(k,\epsilon)&={-\frac{c\hbar v}{q_0}k^2}-\Sigma_3(k,\epsilon),
\end{align}
and
\begin{align}
x(k,\epsilon)&=\frac{1}{X(k,\epsilon)},
\\
y(k,\epsilon)&=-\frac{Y(k,\epsilon)}
{\qty[X(k,\epsilon)+Z(k,\epsilon)]^2-Y(k,\epsilon)^2},
\\
z(k,\epsilon)&=
\frac{Y(k,\epsilon)^2-Z(k,\epsilon)
\qty[X(k,\epsilon)+Z(k,\epsilon)]}
{\qty{\qty[X(k,\epsilon)+Z(k,\epsilon)]^2-Y(k,\epsilon)^2}X(k,\epsilon)}.
\end{align}

Here, we simplify the above expressions for the self-energy into a form more suitable for solving the self-consistent equation. By substituting Eq.~(\ref{green function2}) into Eq.~(\ref{self energy}), we obtain:
\begin{align}
	&
\hat{\Sigma}(\bm{k},\epsilon+is0)
\notag\\ &
= \hat{S}_0\int\frac{k'^2dk'}{(2\pi)^3}n_{\text{i}}\left[V_0^2(k,k')x(k',\epsilon+is0)\right.\nonumber\\
& \hspace{7em}
\left.+(V_0^2(k,k')-V_2^2(k,k'))z(k',\epsilon+is0)\right]\nonumber\\
&\quad
+(\hat{\bm{S}}\cdot\bm{n})\int\frac{k'^2dk'}{(2\pi)^3}n_{\text{i}}V_1^2(k,k')y(k',\epsilon+is0)
\nonumber\\
&\quad
+(\hat{\bm{S}}\cdot\bm{n})^2\int\frac{k'^2dk'}{(2\pi)^3}n_{\text{i}}^2\left(\frac{3}{2}V_2^2(k,k')-\frac{1}{2}V_0^2(k,k')\right)\nonumber\\
&\hspace{10em}\times 
z(k',\epsilon+is0),
\end{align}
with the aid of the useful relations provided in Appendix \ref{integral}. By comparing this with Eq.~(\ref{self energy2}), the self-consistent equation can be decomposed into the following three equations:
\begin{align}
	&
\Sigma_1(k,\epsilon+is0)
 =\int\frac{k'^2dk'}{(2\pi)^3}n_{\text{i}}
 \bigl[V_0^2(k,k')x(k',\epsilon+is0)
\nonumber\\& \hspace{4em} +
(V_0^2(k,k')-V_2^2(k,k'))z(k',\epsilon+is0) \bigr],
\label{self1}
\\ &
\Sigma_2(k,\epsilon+is0)
=\int\frac{k'^2dk'}{(2\pi)^3}n_{\text{i}}V_1^2(k,k')y(k',\epsilon+is0),
\label{self2}
\\&
\Sigma_3(k,\epsilon+is0) = \int\frac{k'^2dk'}{(2\pi)^3}n_{\text{i}}\left(\frac{3}{2}V_2^2(k,k')-\frac{1}{2}V_0^2(k,k')\right)\nonumber\\&
\hspace{12em}
\times z(k',\epsilon+is0).
\label{self3}
\end{align}
By substituting the self-energy into Eq.~(\ref{dos}), the density of states can be expressed as
\begin{align}
&D(\epsilon)
= -\frac{1}{\pi}\Im\int\frac{d\bm{k}}{(2\pi)^3}\left(\frac{1}{X(k,\epsilon+i0)} \right.
\nonumber\\&\quad
+\frac{1}{X(k,\epsilon+i0)+Y(k,\epsilon+i0)+Z(k,\epsilon+i0)}\nonumber\\
&\quad\left.
+\frac{1}{X(k,\epsilon+i0)-Y(k,\epsilon+i0)+Z(k,\epsilon+i0)}\right).
\end{align}

In addition, the Bethe-Salpeter equation can be simplified into a more convenient form. The current vertex $\hat{J}_x(\bm{k},\epsilon,\epsilon')$ is expanded into eight terms as follows: 
\begin{align}
\hat{J}_x(\bm{k},\epsilon,\epsilon')
&=\hat{S}_xJ_0(k,\epsilon,\epsilon')+n_x(\hat{\bm{S}}\cdot\bm{n})^2J_1(k,\epsilon,\epsilon')\nonumber\\
&+n_x(\hat{\bm{S}}\cdot\bm{n})J_2(k,\epsilon,\epsilon')+(\hat{\bm{S}}\cdot\bm{n})^2\hat{S}_xJ_3(k,\epsilon,\epsilon')\nonumber\\
&+\hat{S}_x(\hat{\bm{S}}\cdot\bm{n})^2J_4(k,\epsilon,\epsilon')+(\hat{\bm{S}}\cdot\bm{n})\hat{S}_xJ_5(k,\epsilon,\epsilon')\nonumber\\
&+\hat{S}_x(\hat{\bm{S}}\cdot\bm{n})J_6(k,\epsilon,\epsilon')+n_x\hat{S}_0J_7(k,\epsilon,\epsilon')\label{J},
\end{align}
 by applying Eqs.~(\ref{int1})--(\ref{int2}) to Eq.~(\ref{Bethe}). The matrix-form Bethe-Salpeter equation is thereby reduced to eight separate equations for the expansion coefficients $J_0$ through $J_7$.

From Appendix~\ref{integral}, the Bethe-Salpeter equation is transformed into 
\begin{widetext}
\begin{align}
\left( \begin{array}{c}  J_0\\
 J_1\\
 J_2\\
 J_3\\
 J_4\\
 J_5\\
 J_6\\
 J_7 \end{array}\right)
&=
\left( \begin{array}{c}  1\\
0\\
0\\
0\\
0\\
{{c k}/{q_0}}\\
{{c k}/{q_0}}\\ 
{-{2c k}/{q_0}} \end{array}\right)
\notag\\&\quad
+
\int \displaystyle\frac{k'^2dk'}{(2\pi)^3}n_{\text{i}}\begin{pmatrix}
V_0^2 & 0 & \frac{1}{2}V_0^2-\frac{1}{2}V_2^2 & V_0^2-V_2^2 & V_0^2-V_2^2 & 0 & 0 & 0 \\
0 & \frac{5}{2}V_3^2-\frac{3}{2}V_1^2 & 0 & 0 & 0 & 0 & 0 & 0 \\
0 & 0 & \frac{3}{2}V_2^2-\frac{1}{2}V_0^2 & 0 & 0 & 0 & 0 & 0 \\
0 & 0 & 0 & \frac{3}{2}V_2^2-\frac{1}{2}V_0^2 & 0 & 0 & 0 & 0 \\
0 & 0 & 0 & 0 & \frac{3}{2}V_2^2-\frac{1}{2}V_0^2 & 0 & 0 & 0 \\
0 & \frac{1}{2}V_1^2-\frac{1}{2}V_3^2 & 0 & 0 & 0 & V_1^2 & 0 & 0 \\
0 & \frac{1}{2}V_1^2-\frac{1}{2}V_3^2 & 0 & 0 & 0 & 0 & V_1^2 & 0 \\
0 & V_1^2-V_3^2 & 0 & 0 & 0 & 0 & 0 & V_1^2 \\
\end{pmatrix}\hat{T}
\left( \begin{array}{c}  J_0'\\
 J_1'\\
 J_2'\\
 J_3'\\
 J_4'\\
 J_5'\\
 J_6'\\
 J_7'\end{array}\right)\label{vertex},\nonumber\\
\end{align}
\end{widetext}
where $J_i=J_i(k,\epsilon+is0,\epsilon+is'0),J'_i=J_i(k',\epsilon+is0,\epsilon+is'0), V_i^2=V_i^2(k,k'),x=x(k',\epsilon+is0),x'=x(k',\epsilon+is'0)$ and so on.
The matrix $\hat T$ is defined as
%
%\clearpage
\begin{align}
\hat{T}=\begin{pmatrix}
xx' & 0 & 0 & 0 & 0 & 0 & 0 & 0 \\
T_{01} & T_{11} & T_{21} & T_{31} & T_{41} & T_{51} & T_{61} & T_{71} \\
T_{02} & T_{12} & T_{22} & T_{32} & T_{42} & T_{52} & T_{62} & T_{72} \\
zx' & 0 & 0 & T_{33} & 0 & yx' & 0 & 0 \\
xz' & 0 & 0 & 0 & T_{44} & 0 & xy' & 0 \\
yx' & 0 & 0 & yx' & 0 & T_{55} & 0 & 0 \\
xy' & 0 & 0 & 0 & xy' & 0 & T_{66} & 0 \\
0 & 0  & 0  & 0  & 0  & 0  & 0  & xx' \\
\end{pmatrix}.
\end{align}
Here, the matrix elements $T_{ij}$ in the second row of $\hat{T}$ are expressed as 
\begin{align}
T_{01}&=yz'+zy',
\\
T_{11}&=xx'+xz'+yy'+zx'+zz',
\\
T_{21}&=xy'+yx'+yz'+zy',
\\
T_{31}&=xy'+yz'+zy',
\\
T_{41}&=yx'+yz'+zy',
\\
T_{51}&=xz'+yy'+zz',
\\
T_{61}&=yy'+zx'+zz',
\\
T_{71}&=xz'+yy'+zx'+zz',
\end{align}
$T_{ij}$ in the third row are given by
\begin{align}
T_{02}&=yy'+zz',
\\
T_{12}&=xy'+yx'+yz'+zy',
\\
T_{22}&=xx'+xz'+yy'+zx'+zz',
\\
T_{32}&=xz'+yy'+zz',
\\
T_{42}&=yy'+zx'+zz',
\\
T_{52}&=xy'+yz'+zy',
\\
T_{62}&=yx'+yz'+zy',
\\
T_{72}&=xy'+yx'+yz'+zy',
\end{align}
and the others are given by
\begin{align}
T_{33}&=xx'+zx',
\\
T_{44}&=xx'+xz',
\\
T_{55}&= xx'+zx',
\\
T_{66}&=xx'+xz'.
\end{align}

%By solving these eight component equations [Eq.~(\ref{vertex})], the values of $J_0$ through $J_7$ are determined. 
The expression of conductivity reduces to
\begin{widetext}
\begin{align}
\sigma(\mu)
&= \frac{2\hbar e^2 v^2}{3}\int d\epsilon{\qty(
    -\frac{\partial f}{\partial\epsilon}
    )}\int_0^\infty \frac{k'^2dk'}{(2\pi)^3}\mbox{Re}\left[-\frac{J_0^{++}+J_1^{++}+J_2^{++}+J_3^{++}+J_4^{++}+J_5^{++}+J_6^{++}+J_7^{++}}{(X+Y+Z)^2}\right.\nonumber\\ &
-\frac{J_0^{++}-J_1^{++}+J_2^{++}+J_3^{++}+J_4^{++}-J_5^{++}-J_6^{++}-J_7^{++}}{(X-Y+Z)^2}
-
\frac{q_0+ck}{q_0}
\frac{2J_0^{++}+J_3^{++}+J_4^{++}+J_5^{++}+J_6^{++}}{X(X+Y+Z)}
\nonumber\\ &
-
\frac{q_0-ck}{q_0}
\frac{2J_0^{++}+J_3^{++}+J_4^{++}-J_5^{++}-J_6^{++}}{X(X-Y+Z)}
 -\frac{2c k}{q_0}\frac{-J_7^{++}}{X^{2}}
\notag\\&
+\frac{J_0^{+-}+J_1^{+-}+J_2^{+-}+J_3^{+-}+J_4^{+-}+J_5^{+-}+J_6^{+-}+J_7^{+-}}{|X+Y+Z|^2}
    \nonumber\\ &
+
\frac{J_0^{+-}-J_1^{+-}+J_2^{+-}+J_3^{+-}+J_4^{+-}-J_5^{+-}-J_6^{+-}-J_7^{+-}}{|X-Y+Z|^2}
+
\frac{q_0+ck}{q_0}
\frac{J_0^{+-}+J_4^{+-}+J_6^{+-}}{X(X^*+Y^*+Z^*)}
\nonumber\\ &
+
\frac{q_0-ck}{q_0}
\frac{J_0^{+-}+J_4^{+-}-J_6^{+-}}{X(X^*-Y^*+Z^*)}\left.
+
\frac{q_0+ck}{q_0}
\frac{J_0^{+-}+J_3^{+-}+J_5^{+-}}{X^*(X+Y+Z)}
+
\frac{q_0-ck}{q_0}
\frac{J_0^{+-}+J_3^{+-}-J_5^{+-}}{X^*(X-Y+Z)}+\frac{2c k}{q_0}\frac{-J_7^{+-}}{|X|^2}\right],
\end{align}
where $J_i^{ss'}=J_i(k',\epsilon+is0,\epsilon+is'0)$, $X=X(k',\epsilon+i0)$, and so on.
\end{widetext}

\bibliography{ref}
\clearpage

\end{document}